# TEL AVIV UNIVERSITY

The Iby and Aladar Fleischman Faculty of Engineering

The Zandman-Slaner School of Graduate Studies

# POTENTIALS AND LIMITS OF SUPER-RESOLUTION ALGORITHMS AND SIGNAL RECONSTRUCTION FROM SPARSE DATA

A thesis submitted towards the degree of
Master of Science in Electrical Engineering

by

# Gil Shabat

June   2008

# TEL AVIV UNIVERSITY

The Iby and Aladar Fleischman Faculty of Engineering

The Zandman-Slaner School of Graduate Studies

# POTENTIALS AND LIMITS OF SUPER-RESOLUTION ALGORITHMS AND SIGNAL RECONSTRUCTION FROM SPARSE DATA

A thesis submitted towards the degree of

Master of Science in Electrical Engineering

by

# Gil Shabat

This research was carried out in the School of Electrical Engineering

under the supervision of Prof. Leonid Yaroslavsky

June   2008

# Acknowledgements

I would first like to express my deep gratitude to my supervisor, Professor Leonid Yaroslavsky. His wide knowledge, patience and personal guidance throughout this work have been a great value for me. I feel very fortunate for having a supervisor like you.

I wish to express my sincere gratitude to Barak Fishbain, for constructive and interesting discussions and for valuable advices and friendly help.

# ABSTRACT

A common distortion in videos is image instability in the form of chaotic global and local displacements of image frames caused by camera instability, fluctuations in the refraction index of the light propagation media and similar factors. Such videos that very frequently present moving objects on a stable background contain tremendous redundancy that potentially can be used for image stabilization and perfecting provided reliable separation of stable background from true moving objects. Recently, it was proposed to use this redundancy for resolution enhancement of video through elastic registration, with sub-pixel accuracy, of segments of video frames which represent stable scenes.

The present work is aimed at studying the potentials and limitations of such a resolution enhancement. The work consists of two parts. In the first part we investigate, by means of computer simulation, the influence on the degree of the achievable resolution enhancement such imaging parameters as the camera fill factor, of intensity of fluctuations of pixel displacements and of the number of image frames. The essential part of the process of resolution enhancement is signal reconstruction from sparse data accumulated from the set of randomly displaced image frames. Therefore in the second part of the work we address the theory of discrete signal reconstruction from sparse data. We introduce the discrete sampling theorem and show that given finite number of image samples which is lower than that defined by the required sampling grid one can reconstruct a band-limited, in terms of a selected basis, approximation of the signal with minimal mean squared error. We also analyze limitation imposed by different bases to the positions of available sparse signal samples to secure optimal restoration and show that low pass band-limited in DFT basis functions can be precisely reconstructed from their arbitrary placed sparse samples. These results are then extended to image reconstruction from limited number of projections or from projections with partly lost samples.

# TABLE OF CONTENTS

# LIST OF SYMBOLS

| | |
|---|---|
| LR | Low Resolution |
| HR | High Resolution |
| SR | Super Resolution |
| 1D | One Dimensional |
| 2D | Two Dimensional |
| BW | Band Width |
| CCD | Charge Coupled Device |
| STD | Standard Deviation |
| MSE | Mean Square Error |
| RMS | Root Mean Square |
| RGB | Red Green Blue |
| DCT | Discrete Cosine Transform |
| DFT | Discrete Fourier Transform |
| DWT | Discrete Wavelet Transform |
| FFT | Fast Fourier Transform |
| SDFT | Shifted Discrete Fourier Transform |
| PPFT | Pseudo Polar Fourier Transform |
| D4 | Daubechies Wavelet of Type 4 |
| CT | Computerized Tomography |

# TABLE OF FIGURES

# INTRODUCTION

Super-Resolution (SR) is a general term for a set of methods for increasing image (or video) resolution. All SR techniques are based on the same idea: using information from several images, to create an up-sampled one. These methods try to extract details from each given frame and fuse all the accumulated data into a single new frame. The captured low-resolution (LR) frames, which are slightly different images of the same object or scene are used as the input for the SR algorithm. If the frames are exactly the same, no information is added, and the SR image cannot be resolved. Therefore, most super-resolution techniques are based on the fact that no real observation platform can be absolutely stationary. There are always some micro-movements during the video data acquisition stage. Consequent frames that differ only due to these small movements of the image plane can be combined in order to generate a new image with better spatial resolution. Generally, SR techniques can be divided into three main stages: The first is determination with sub-pixel accuracy of the location of every pixel in each LR frame; this can be done by applying an optical flow or a motion estimation algorithm to the input LR video. The second is accumulation of the data into a new image to obtain an up-sampled image. The third stage is interpolating the new image to obtain a complete image and to remove aliasing. Super-resolution principles and general multi-channel image recovery are detailed in the works of S. Srinivasan and R. Chellappa [1], Galatsanos, Wernick and Katsaggelos [2], R.R. Schultz [3] and T. J. Schultz [4]. Some researchers formulated solution to global motion scenarios, usually from an application perspective of a non-stationary camera or zooming.

Irani and Peleg [5], suggested a super-resolution approach based on registration of several images and projecting their shifted samples into a new "initial guess" to create an SR image. SR for infrared image using global motion is presented in [6], where the forward looking infrared (FLIR) array detector is located on a moving object such as a car or an aircraft. Image registration is used to place all the samples on a new grid and to produce super resolved image. Other SR methods which rely on global motion (non-stationary camera) and image registration can be found in [7, 8, 9]. A different kind of SR obtained from multiple images, this time through zooming instead of motion was suggested by Li [10]. In his work, the input LR frames are zoomed images of the same object. Since the number of samples remains the same in every zoom, but the area captured by the camera is smaller, the samples are taken in different positions. Fusing all those samples to a new denser grid creates a "synthetic zoom".

The first part of this thesis, concentrates on SR from turbulent videos. Atmospheric turbulence is a major cause for image distortion in long distance observation systems. Those turbulences cause spatial random fluctuations in the refraction index of the atmosphere [11].At the result, light from each of the points in the scene acquires slightly different tilts and low order aberrations, causing the images of these points to be randomly dislocated from their correct geometrical positions. In acquiring these images by video cameras, the image sampling grid defined by the video camera sensor can be considered to be chaotically moving over a stationary image scene. Therefore, in turbulence-corrupted videos, consequent frames of a stable scene differ only due to atmospheric turbulence-induced local displacements between images. This phenomenon allows generating images of such scenes with a larger number of samples than that provided by the camera if consecutive image frames are combined by means of their appropriate re-sampling. Resolution enhancement from turbulent videos was investigated by Charnotskii ([12]) who showed preliminary feasibility for super-resolution in physical experiments taken in a laboratory. These experiments, based on taking photos of an object through turbulent medium, showed a certain potential for super-resolution by using multiple turbulence distorted frames. Fraser *el al.* ([13],[14]) have suggested a similar method for increasing image resolution by using turbulence distorted videos. In their work, the SR image is obtained by first creating a relatively blurred reference frame of temporal average over the LR distorted frames, and then calculating the local shifts between each LR frame and the reference frame using local correlation or optical flow. Then, the reference frame is re-calculated using the corrected frames, and the process is repeated iteratively. This suggested algorithm requires high system resources and is not suitable for real time applications. Also, no interpolation method is used to remove all the aliasing and convergence is not discussed. While the turbulent motion was previously dealt, none of the previous researchers have suggested using turbulent motion for super-resolution applications based on real-time digital image processing. The implementation discussed in this work is based on creating a reference frame by calculating the median, finding local shifts by using correlation or optical flow based methods and a good interpolation method for removing aliasing and image restoration from sparse data. Furthermore, this research tries to answer the question "What can be achieved?" or "What are *the potentials and limitations* of super-resolution from turbulent videos?" for different scenarios and under certain constraints, such as given number of frames, camera's PSF, etc. ([15],[16])

The second part of this work deals with reconstruction of discrete signals from sparse data. A lot of work has been done for the case of reconstructing non-uniformly sampled band-limited continuous signals. Today however, when digital computers play a major role in modern science and engineering, signals which are continuous in their physical nature are stored digitally in computers by their discrete representation, the importance of sampling and reconstructing discrete signals such as audio, image and other signals has increased dramatically. The problem of signal and image reconstruction is usually attacked by means of interpolation methods. Though many interpolation methods exist for continuous and discrete signals based on diverse ideas ([17],[18],[19],[20],[21]) none of them take into account the sampling basis of the signal and therefore, are not optimal. Shepard's method ([17]) was one of the first methods for interpolating a 2D irregularly sampled signal. The idea behind Shepard's method is simply interpolating in every missing point by weighting the known samples inversely to their distance, so that close samples will have bigger weights. The method's disadvantage is that the interpolated function doesn't look like the original near the known points. Another well known classical method is the Hardy's multiquadrics [18]: In this method, the interpolating function is a linear combination of radial basis functions, where the coefficients are determined by the interpolation conditions. Splines are also a useful tool in interpolation and approximation. In [19], Wolberg *et al*, describe an algorithm for interpolation and approximation based on multilevel B-Splines. The idea is approximating a function from its sparse data using control points lattice. The control points are chosen so that the value at the known points will not change (if possible). After coarse approximation is made, a denser grid of control points is created to approximate the error, and the process continues iteratively. Another interesting interpolation method, was suggested in a paper by Baraniuk and Choi ([20]). In this paper, the interpolated function belongs to some wavelet domain and therefore can be described by their coefficients. Since the function is given only by its sparse data, the problem reduces to finding the best set of wavelet coefficients that describes the function. Since there are an infinite number of possibilities, the chosen solution is the one that minimizes the Besov norm, which is a smoothness criterion (or equivalently: band limitation). The smoothness of the interpolating function is taken as an a-priori knowledge (a parameter in the Besov norm). Similar and interesting, spline-based approach was suggested by Margolis and Eldar ([21]). The idea is based on interpolation with non-uniform splines in which works as a basis function of the interpolated signal. In this paper, an interpolation is performed to a CT image, under the assumption that its Fourier transform has finite support (band-limited on Fourier domain). In this thesis, an optimal signal recovery in RMS sense is presented by taking into account the basis functions that created the sampled signal as an a-priori knowledge, along with an elaborate analysis of sampling and reconstructing band-limited discrete signals is presented, where the band-limitation of the discrete signals is not restricted only to the Fourier domain, but extended to any invertible transform such as wavelet and Radon. Analysis for the 1D and 2D signals on the Fourier, Haar wavelet, D4-wavelet and Walsh domains is presented and example for exact reconstruction of medical CT images on Radon domain is introduced as a practical application of the theory. The first chapter of this thesis gives some background about work that has been recently done, about stabilization and super-resolution of turbulent videos. The second chapter describes the computer model that was built for investigating the potentials and limitations that turbulent videos possess for producing super-resolution images from distortions and explains what are the parameters being investigated and their influence on the quality of the obtained image. Chapter 3 shows the results from the computer model for different parameters. Chapter 4 introduces the theory of discrete sampling and develops conditions and tools for perfect reconstructions. Chapter 5 shows an example for the theory of chapter 4 for reconstructing of medical CT images from sparse data using an exact inverse Radon transform. Chapter 6 summarizes the entire work.

# 1   BACKGROUND AND MOTIVATION FOR THIS WORK

In many real life scenarios, when acquiring images, it is common to have image distortions in form of global and local displacements caused by camera instability, fluctuations in the refraction index of the light propagation media and similar factors. Such videos that very frequently present moving objects on a stable background contain tremendous redundancy that potentially can be used for image stabilization and perfecting, provided reliable separation of stable background from true moving objects. An example for a distortion that causes local displacements to a stable background is atmospheric turbulence, which is quite common in long range observation systems. In a paper by Fishbain *et al* [22], it was proposed to use this redundancy and instabilities for resolution enhancement of videos through elastic registration with sub-pixel accuracy. This method of stabilization and resolution enhancement consists of two basic steps. The first step is separating the actual moving objects from the stable background. This is done by applying a motion estimation (such as Lucas-Kanade, Horn-Schunck, MPEG4, etc.) algorithm and relying on the fact that image instabilities caused by turbid media are usually small and stochastic, where actual movements are more consistent. Once the motion vectors were extracted with respect to a reference frame (temporal median), it is possible to distinguish between the actual movements and the stochastic shifts caused from the instabilities of the turbid media. The second step is using the small shifts that were determined to sub-pixel accuracy for resolution enhancement by accumulating pixels to a new denser grid and leaving the actual moving object untouched. A result of the SR algorithm described in this work is shown in Figure 1-1. In this experiment, color tiles were laid inside a water tub, where the turbulent affect is caused by the filling water. The light propagates through the water turbulent medium, was captured by a CCD camera. The input for the SR algorithm was 300 LR frames, with average shift of 2.5 pixels and STD of 1.3 pixels.

In order to analyze the potentials (and limitations) this method of resolution enhancement possesses, a computer model was built. The computer model allowed us to investigate the influence different parameters had on the performance of the super-resolution (SR) algorithm and answer questions such as "How does the fill factor of the CCD affect the SR image?" or "How does the interpolation method applied at the end of the SR process influence the final image obtained?". All these questions and others can be answered by using computer simulation. The computer simulation that was built, simulates atmospheric turbulence, and therefore gives complete control over the turbulent distorted video used as an input for the SR algorithm, and also provides complete control over the different parameters of the SR algorithm itself. Since a major part of the SR process is reconstructing an image from its sparse samples, the second part of this work is dedicated to the problem of interpolating discrete signals. Most interpolation methods used today and mentioned above, in the introduction, were developed on the theoretical foundations of continuous signals but are being applied to discrete signals when the signal is stored in a digital computer. In this work, a theory for signal reconstruction aimed to discrete signals is introduced for both direct and iterative discrete signal reconstruction, optimal in the MSE sense.

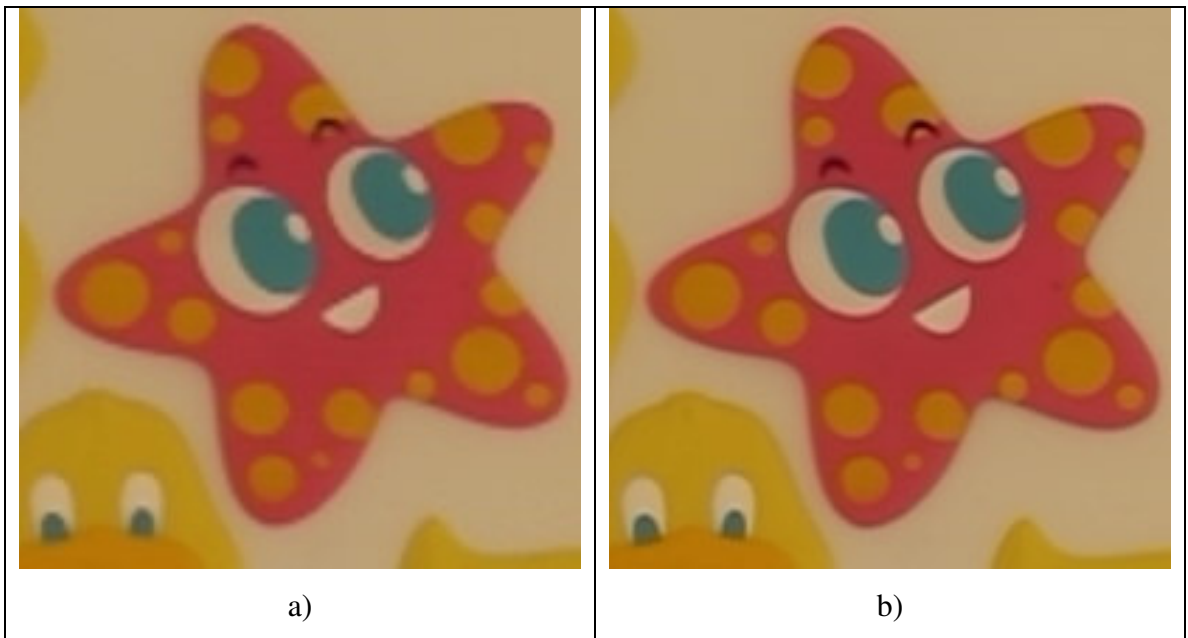

a)          b)

**Figure 1-1: Results of SR algorithm, a) Stabilized LR image b) SR image**

# 2 COMPUTER MODEL

In order to understand the potentials of SR from turbulent distorted video, it was required to create a computer model for investigating the influence of the different parameters on the performance of the resolution enhancement that can be achieved with the algorithm. Using the computer model, it was possible to generate a variety of turbulent videos for different scenarios. Once the set of LR frames (the LR turbulent video) was produced, it was possible to apply the SR algorithm to obtain results and evaluate them in the light of the parameters that were used. In this work, 4 parameters were taken into account: Fill factor (which can be extended to include the general Point Spread Function (PSF) of the optics in the camera), Number of frames in the LR video, turbulence intensity and number of iterations in interpolation process.

## 2.1 Computer Model Parameters

The following parameters were investigated in this research:

### 2.1.1 Camera Fill Factor

The camera fill factor is the ratio of the active detection area (the size of the light sensitive photodiode) to the inter pixel distance as shown in Figure 2-1, where the gray area marks the active detection area and white area along with the gray is the entire sensor. Because of the electronics around the pixels, fill factor value is smaller than or equal to 1 (the ratio between the gray area to the entire sensor area). Camera photo detectors introduce low pass filtering to the images captured by the camera. Large fill factor means better light energy efficiency of photo detectors and higher degree of low pass filtering, which causes a loss of image high spatial frequencies. Figure 2-2 illustrates frequency responses of photo detectors with different fill factors. The resolution of images acquired by cameras is ultimately limited by this low pass filtering. Super-resolution methods allow eliminating aliasing effects due to image sampling but not the image low pass filtering by camera photo detectors. The latter can be, at least partially, compensated by means of deblurring through aperture correction. Therefore, super-resolution methods are potentially more efficient for images acquired with cameras with small fill factor.

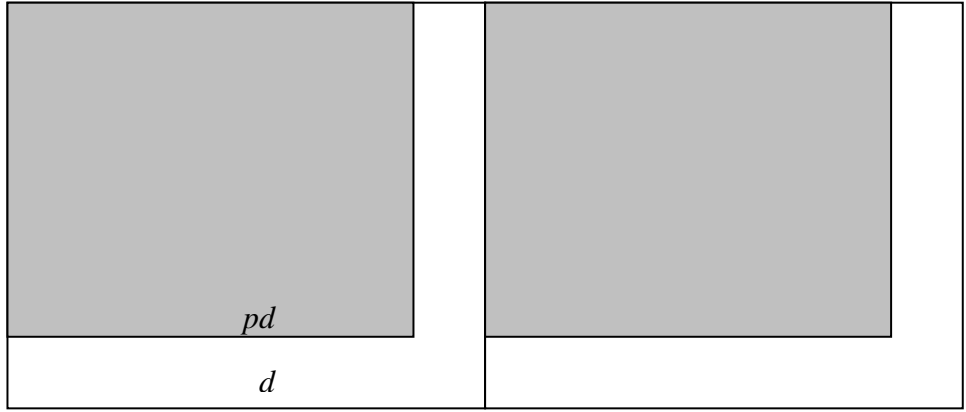

**Figure 2-1: Illustration of the fill factor and the inter pixel distance**

In rectangular grids, such as in Figure 2-1, the fill factor frequency response is given by a continuous sinc function (Fourier transform of a rectangular): $\mathrm{sinc}(\frac{pd\omega}{d}) = \mathrm{sinc}(p\omega)$, where $d$ is the width of the sensor, and $p$ is the fill factor $0 < p \leq 1$ as shown in Figure 2-2.

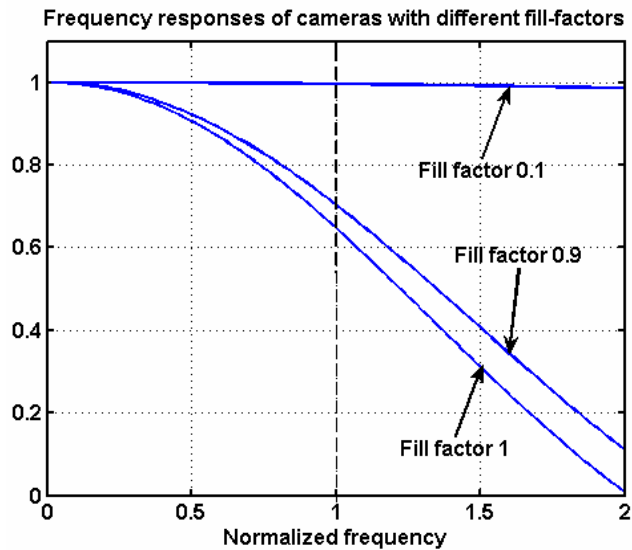

**Figure 2-2: Camera frequency responses for large and small fill factors. Frequency axis is normalized to the width of the camera base band**

On the computer model, the fill factor was simulated by upsampling the image by M, using discrete sinc interpolation to the image, and then applying a 2D low-pass filter. In the case of rectangular sensor as discussed here, the filter that was applied here was a 2D local mean of size $K \times K$ where $K/M$ is the required fill factor:

$$h = \frac{1}{K^2} \begin{pmatrix} 1 & 1 & \ldots & 1 \\ 1 & & & 1 \\ \vdots & & & \vdots \\ 1 & 1 & \ldots & 1 \end{pmatrix}_{K \times K} \quad \text{(2-1)}$$

In case it will be needed to use the model when taking into consideration the PSF of the lens then the filter $h * PSF$ should be applied, where * stands for convolution.

### 2.1.2 Turbulence Intensity

Pixel displacements due to atmospheric turbulence are chaotic and therefore can be characterized only statistically. In our study, the intensity of the turbulence was specified by the standard deviation of the motion vector length. Weak turbulence with low standard deviation of the motion vector length causes small shifts. In this case, the low-resolution (LR) video will virtually appear static and the frames will be nearly identical. Intensive turbulence with standard deviation by the order of the inter-pixel distance may cause very substantial aliasing, which might be difficult to measure and compensate. One can expect that standard deviation of motion vector length of about 0.5 is potentially most suitable from the point of view of the super-resolution process.

### 2.1.3 The number of processed low resolution frames

The quality of the resolution-enhanced frames depends on the amount of data used for their formation. More frames can potentially produce better quality. However, when the number of frames is becoming large enough, having more frames will not necessarily supply more (or significantly more) new pixels, while it will require more processing time.

## 2.1.4 The number of iterations in the process of re-interpolation

Once the motion vectors for each available low resolution frame are known, pixels in the sub-sampled reference frame are replaced with known pixels from those frames. As a result, an up-sampled reference frame that contains pixels (samples) from all the low resolution images is obtained. At this stage it is required finally to perform image re-interpolation to remove aliasing and to generate the best approximation to the image from the given set of pixels. This is achieved by the iterative interpolation algorithm, which converges to the best band limited approximation of the image. This algorithm will be thoroughly analyzed and discussed in the second part that deals with signal reconstruction from sparse data. More iterations means that the final image will be closer to the best possible approximation within a given bandwidth. However, iterations consume time, and therefore a compromise should be sought between the resulting image quality and computation time.

## 2.2 Creating a Low-Resolution Turbulent Video

The input parameters for the simulation were camera fill factor and the frame-wise pixel translation maps for simulating the turbulence effect. The realizations of the motion vector maps were generated in the form of X- and Y-shift arrays of pseudo-random Gaussian random numbers with a given standard deviation as appears in Figure 2-4. For each realization of the pixel translation map, a corresponding low-resolution frame was produced by means of down-sampling an up-sampled low pass filtered high-resolution image according to the sampling grid specified by the corresponding motion vector map.

Formally, the following steps are applied to the HR test image in order to produce a LR turbulent frame. Step 3 of the process is repeated for every displacement map which corresponds to a LR frame:

1. Take the HR input test image, up-sample it by M using discrete sinc interpolation (DFT or DCT)

2. Apply low pass filter, as described in 2.1.1 to simulate the influence of the fill factor (and camera optics' PSF if needed). The image at this point is marked by $I$ and is M times larger than the original HR image, because of the discrete sinc-interpolation.

3. Let U, V be the (sub-pixel) displacement maps along X and Y respectively. Create a new image $\tilde{I}_k[m,n] = I[N(m-1)+M \times U(m,n), N(n-1)+M \times V(m,n)]$

$\tilde{I}_k$ is an output LR turbulent frame, which is N times smaller than $I$ and is created by irregularly downs-sampling of the image $I$ to simulate the turbulent affect.

The up-sampling factor used in the model was M=10, and the down-sampling factor was N=20. Once low resolution frames were obtained, they were used as an input sequences for the super-resolution algorithm shown in flow diagram of Figure 2-3. Figure 2-5 shows a HR input image, an input motion field and a LR turbulent output image. Figure 2-6 Shows HR and LR Text images.

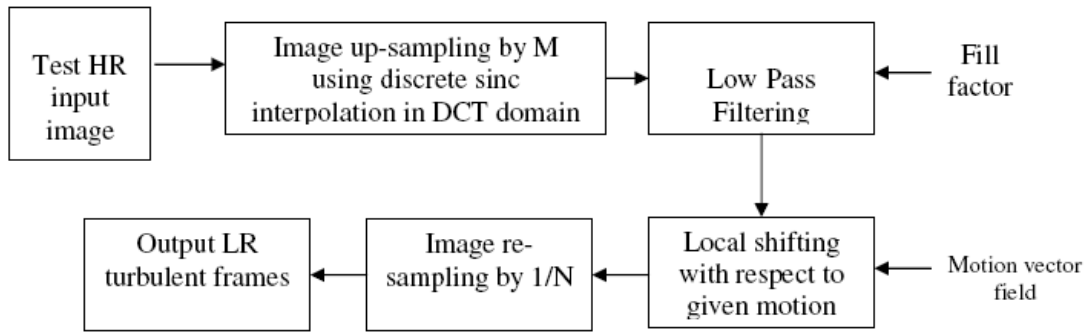

**Figure 2-3: Flow diagram of the computer model**

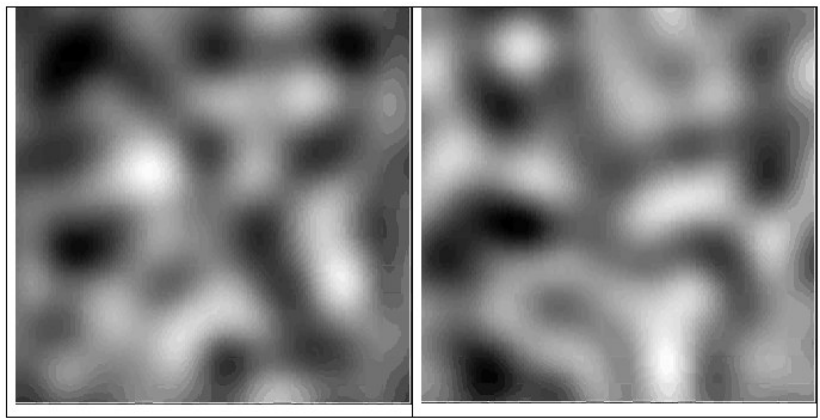

**Figure 2-4: Examples of spatially correlated motion vectors for the X and Y axes**

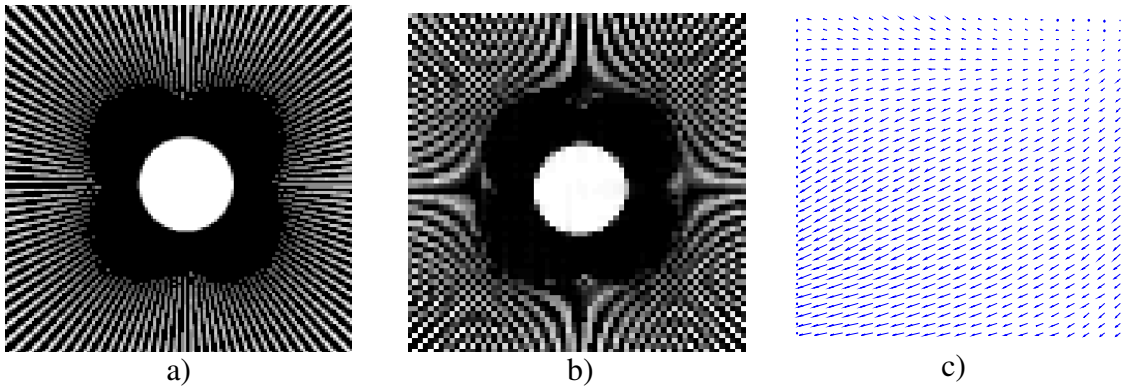

a)　　　　　　　　　　　　b)　　　　　　　　　　　　c)

**Figure 2-5: a) Original high resolution test image, b) a turbulent low resolution frame (output), c) a sample of a motion vector map.**

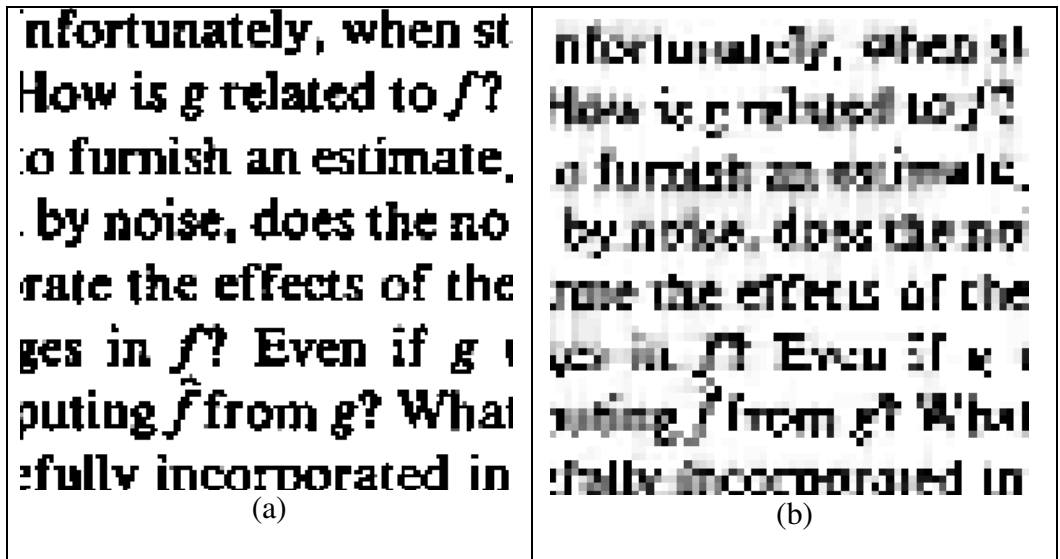

(a)　　　　　　　　　　　　(b)

**Figure 2-6: (a) Original HR Text Image (b) Turbulent LR Text image**

## 2.3　Creating a Super-Resolution Image From the LR Video

Once the LR frames were obtained using the first part of the model described above, it is now possible to take those LR images and produce a resolution enhanced image, using the super-resolution algorithm whose block diagram described in Figure 2-8. For each current frame of the turbulent video, inputs of the process are: a corresponding reference frame, which represents an estimate of the stable scene, and the current frame displacement map. The latter serves for placing pixels of the current frame, according to their positions determined by the displacement map, into the reference frame. This is implemented by means of its corresponding up-sampling to match the sub-pixel accuracy of the displacement map. As a result, output stabilized and enhanced in its resolution frame is accumulated. In this accumulation process it may happen that several pixels of different frames are to be placed in the same location in the output enhanced frame. In order to make best use of all of them, these pixels must be averaged. For this averaging, the median of those pixels is computed in order to avoid the influence of outliers that may appear due to possible errors in the displacement map.

After all available input frames are used in this way, the enhanced and up-sampled output frame contains, in positions where there were substitutions from input frames, accumulated pixels of the input frames and, in positions where there were no substitutions, interpolated pixels of the reference frame. Substituted pixels introduce to the output frame high frequencies outside the base-band defined by the original sampling rate of the input frames. Those frequencies were lost in the input frames due to the sampling aliasing effects. Interpolated pixels that were not substituted do not contain frequencies outside the base-band. In order to finalize the processing and take full advantage of the super-resolution provided by the substituted pixels, the iterative re-interpolation algorithm can be used akin to the Papoulis-Gerchberg algorithm [23,24]. Once iterations are stopped, the output-stabilized and resolution-enhanced image obtained in the previous step is down-sampled to the sampling rate determined by selected enhanced bandwidth and then it can be subjected, when needed, to additional processing aimed at camera aperture correction and, if necessary, denoising.

Formally, the algorithm for establishing SR image from a set of given LR turbulent frames is the following (input: LR video, scaling factor N, up-sampling factor M, displacement maps U,V for each axis; output: SR image):

1. Calculate reference frame from the LR video by temporal median *ref_frame = median(LR_video, 3);*
2. Let [SzX, SzY]=size(LR_frame);

3. Up-sample the reference frame by *N* using discrete sinc interpolation on DFT/DCT domain. Let denote this frame by *SRBaseFrame*.
4. Initialize *SRframe* and *CountMAT* to be a matrix of zeros with size(*SRBaseFrame*)
5. For every LR frame $\tilde{I}_k$ do:
    a. Up-sample the k-th LR frame by *N* using discrete sinc interpolation.
    b. For every pixel (m,n) in $\tilde{I}_k$ do: (Accumulation stage)
        i. Define:
        $$x = \min(N \times szX, (\max(1, ((m-1) \times N + round(N \times U(m,n))))));$$
        ii. Define:
        $$y = \min(N \times szY, (\max(1, ((n-1) \times N + round(N \times V(m,n))))));$$
        iii. *SRframe[x,y]=SRframe[x,y]* + $\tilde{I}_k[m,n]$
        iv. *CountMAT[x,y]=CountMAT[x,y]+1;*
6. Replace all zeros in *CountMAT* by ones.
7. Perform element wise division so that *SRframe* ← *SRframe / CountMAT*
8. *SRframe* is a sparse matrix. Replace all zero values in *SRframe* with values of *SRBaseFrame*
9. Apply Gerchberg-Papoulis interpolation to *SRframe* with fractional bandwidth of 1/M.

The interpolation is performed to fractional baseband of 1/M since this is the bandwidth of the original HR image after applying discrete sinc interpolation with factor M, so this can be taken as an a-priori information for image bandwidth. The outcome of the interpolation stage is the best bandlimited approximation within the original HR image baseband that can be found for the samples accumulated from the LR turbulent video. The SR image is obtained when interpolation ends. Though in the iterative interpolation process, only the pixels that were accumulated from the LR frames are forced to be in their position, replacing the unknown pixels location by pixels from the interpolated reference frame is used as initial guess, to accelerate the convergence of the iterative interpolation. Figure 2-7 shows an LR image, and SR images before and after interpolation.

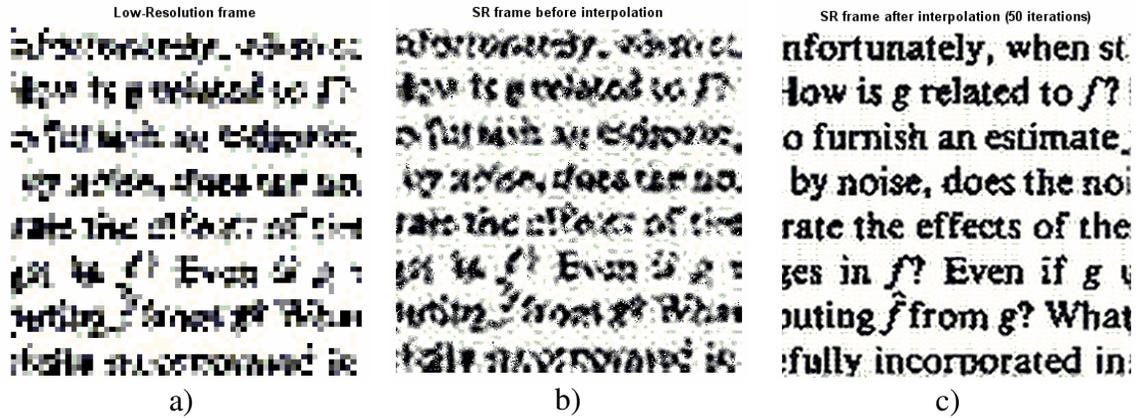

**Figure 2-7:** Iterative image interpolation in the super-resolution process: a) – a low resolution frame; b) Image fused (accumulated) from 50 frames (before interpolation); c) – a result of iterative interpolation of image after 50 iterations.

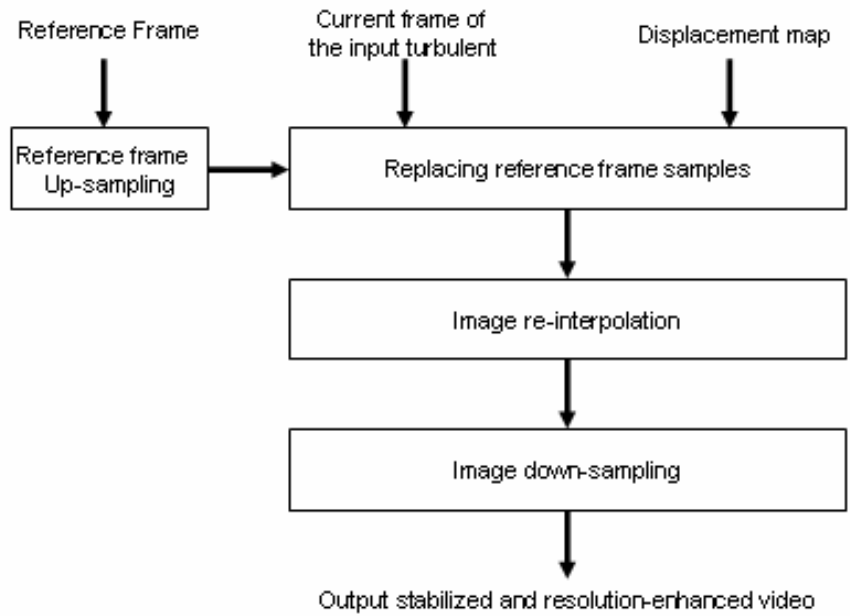

**Figure 2-8: Flow diagram of the process of generation of stabilized frames with super-resolution**

# 3   RESULTS

This part of the thesis shows results that were obtained from the super-resolution algorithm after applying it to simulated turbulent videos which tells us about the potential and limitations of the algorithm. When extracting SR images from real videos, the motion vectors can be found by optical flow or motion estimation methods.

## 3.1   Camera Fill Factor

As it was already mentioned, camera fill factor determines the degree of low pass filtering of images acquired by the camera. Figure 3-1 illustrates results of generating super-resolved images from sequences of turbulence distorted low-resolution images acquired with cameras with fill-factors 0.05, 0.5 and 0.95. In all cases, the number of low-resolution frames used was 30, the standard deviation of the motion vector length was 0.5 pixels and 50 iterations were used for re-interpolation. It can clearly be seen from the figure, that cameras with small fill factor produce better results. Spectra of the corresponding images shown in Figure 3-1, d) through f), also demonstrate that images acquired with larger fill factor have less high frequencies.

## 3.2   Turbulence intensity

Results of studying influence of turbulence intensity on the efficiency of image resolution recovery through the super-resolution process are illustrated in Figure 3-2 and in Figure 3-3. Super-resolved images shown in Figure 3-2 were obtained from 30 low resolution frames, the camera fill factor was 0.05 and 100 iterations were used in the interpolation. One can see from these images that turbulence with standard deviation of motion vector lengths of about 0.5 inter-pixel distances creates a sort of optimal conditions for image resolution recovery.

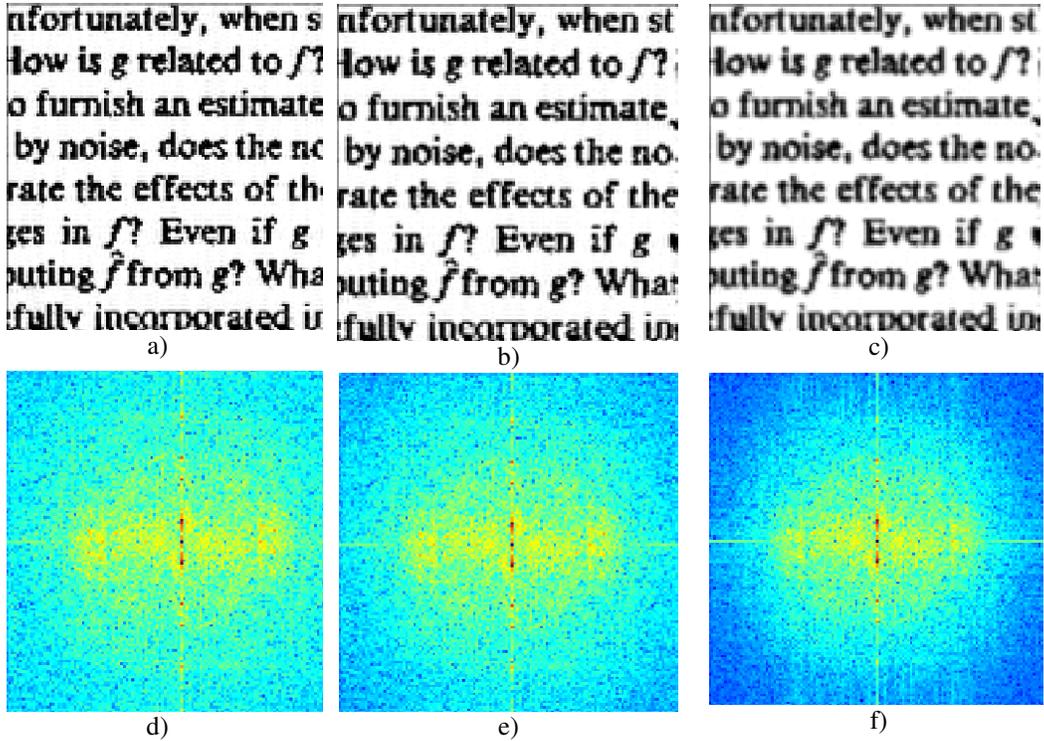

**Figure 3-1–** Super-resolved images obtained from low resolution images acquired by cameras with different fill factors: a) - fill factor 0.05; b) - fill factor 0.5; c) - fill factor 0.95. Figures d)-f) show corresponding image spectra intensities displayed in pseudo colors.

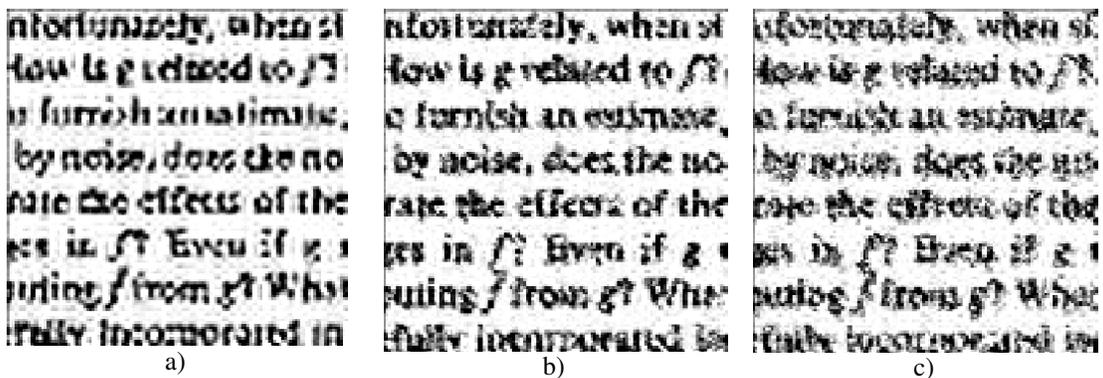

**Figure 3-2 –** Super-resolution results obtained from low resolution images distorted by atmospheric turbulence with different intensity: (a) standard deviation (STD) of motion vector length 0.1 of inter-pixel distance; (b) STD 0.4 of inter-pixel distance; (c) STD of 0.8 of inter-pixel distance.

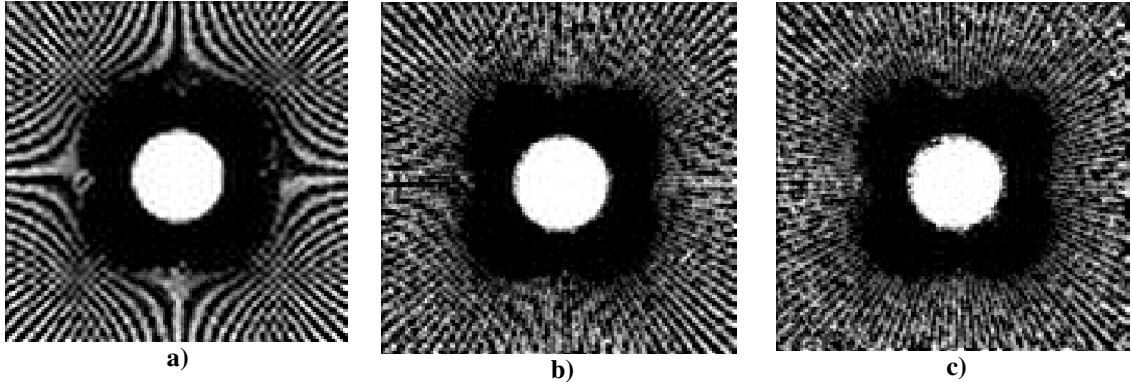

a)    b)    c)

**Figure 3-3:** Super-resolution results obtained from low resolution images distorted by atmospheric turbulence with different intensity: d) standard deviation (STD) of motion vector length 0.1 of inter-pixel distance; e) STD 0.45 of inter-pixel distance; f) STD 0.9 of inter-pixel distance.

## 3.3 The number of processed frames

Obviously, the number of processed low-resolution frames directly affects the super-resolution performance as more frames provide more additional samples to form denser sample grid. The question is how many frames are needed to enable resolution improvement for a given turbulence intensity? Ideally, to obtain two times higher resolution one needs to supply 3 additional samples for each initial low resolution sample which means 3 additional frames for each low resolution frame. Our simulation, however, has shown that in reality the number of additional frames must be much larger. This finding is illustrated in Figure 3-4 and Figure 3-5 for two test images. In these experiments, camera fill factor was 0.05, standard deviation of motion vector length was 0.5 and 50 interpolation iterations were used.

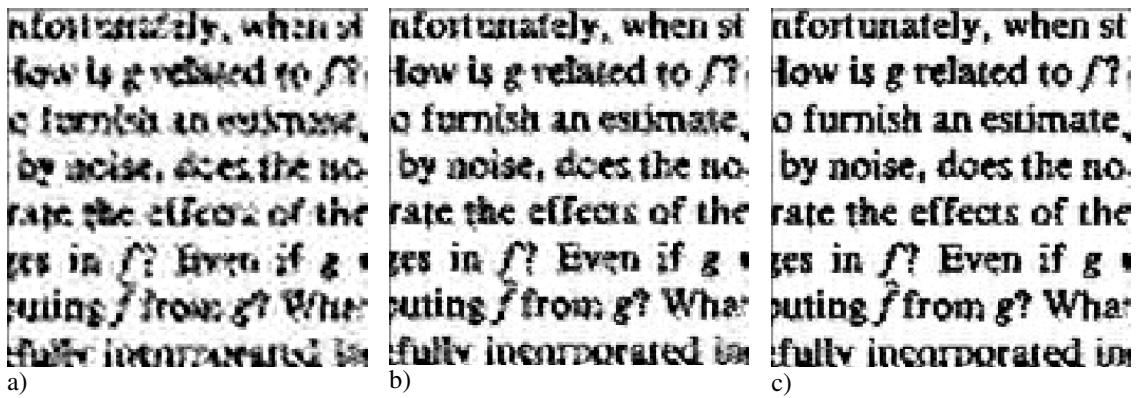

a)    b)    c)

**Figure 3-4: a) – c).** Results of image resolution recovery from 5, 15, and 30 low resolution turbulent images, correspondingly

## 3.4 The number of iterations in the process of re-interpolation

Image re-interpolation is the final stage of the super-resolution process aimed at recovery of those samples in the dense sampling grid that were not obtained from the accumulated low resolution frames. As it was mentioned, it is implemented through an iterative interpolation algorithm, which converges to the best band limited approximation of the image. Figure 3-6 a) – c) show how the number of iterations influences the quality of final super-resolved image. In this experiment, camera fill factor was 0.05, standard deviation of vector motion lengths was 0.5 of inter-pixel distances and 30 low resolution frames were used. Figure 3-7 illustrates the iteration process. It shows a typical dependence of the energy of the difference between subsequent images in course of iterations from the number of interpolation iterations. From these figures one can see that the number of iterations is a quite critical parameter of the restoration process and that, to achieve a good restoration quality, one needs about 100 iterations. Figure 3-7 shows the convergence of the process.

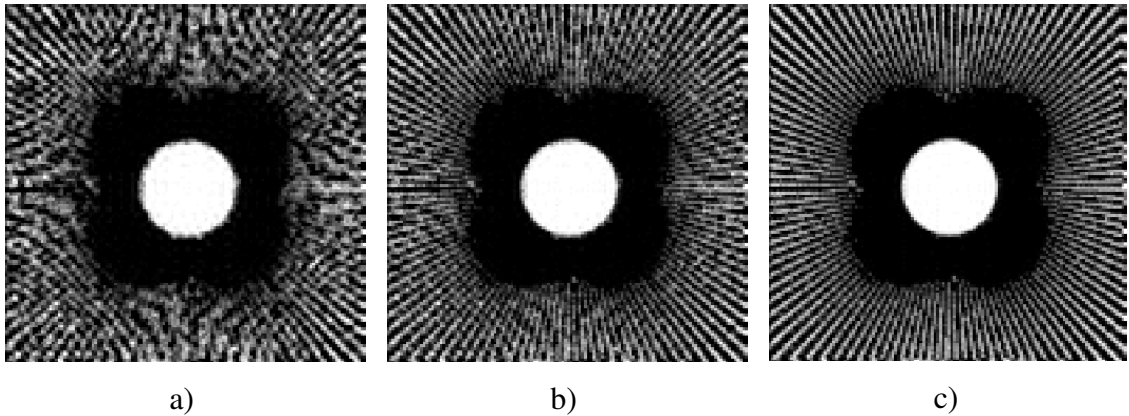

a)                                             b)                                             c)

**Figure 3-5: a) – c). Results of image resolution recovery from 5, 15, and 30 low-resolution turbulent images, correspondingly**

From the results, it can be seen that distortion caused by atmospheric turbulence can be used to increase image resolution beyond camera's limitation. In order to obtain good results, it is required to have some turbulent motion of the scene, to have some sub-pixel re-sampling. Loosely speaking, the more frames in the given video the more information is given and therefore better results can be obtained. However, at some stage more frames will not supply new information and therefore the SR results will reach saturation.

Several dozens of low-correlated frames are required to have significant results. Another limitation over the SR performance is the fill factor of the camera. Large fill factor causes to a loss of high frequencies. Those frequencies are lost and cannot be restored. A good interpolation method is also important, to remove aliasing of the sampling, and to achieve the best approximation to the original HR image from the given samples.

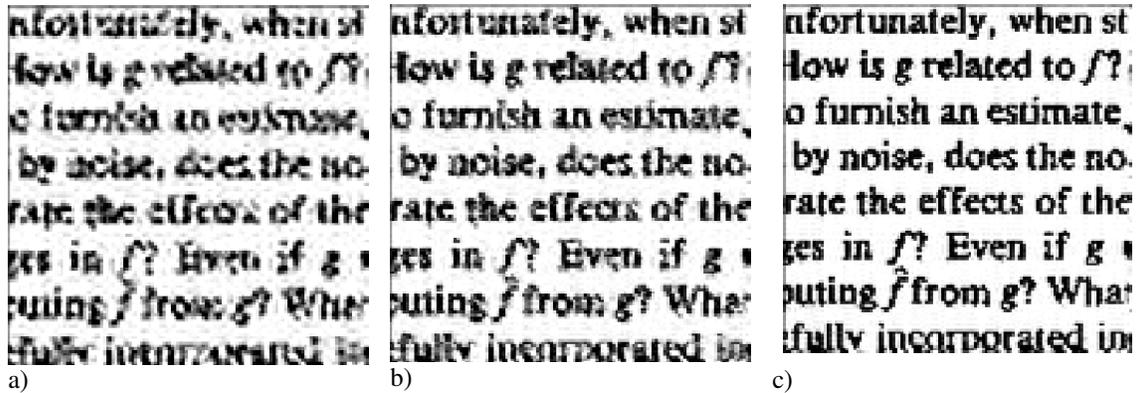

a)  b)  c)

**Figure 3-6: a)-c) SR image with 5, 20 and 100 iterations, respectively. Picture does not fit page margins**

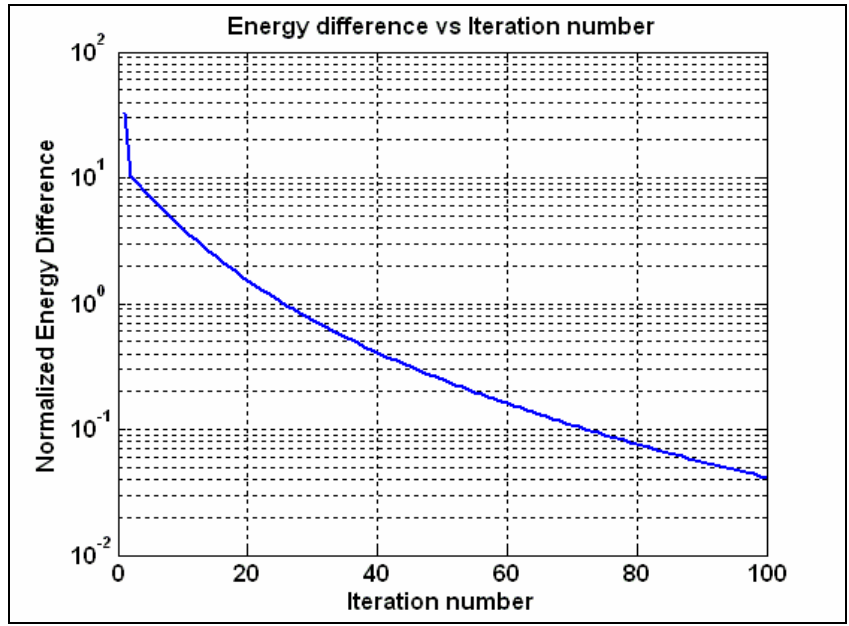

**Figure 3-7: Energy of the difference between subsequent images in course of iterations as a function of the number of interpolation iterations**

# 4  SIGNAL RECONSTRUCTION FROM SPARSE DATA

The last step in the creation of the SR image is reconstruction of the image from the samples that were accumulated on the SR grid. However, this is just a special case of the more general problem, which arises frequently in many fields such as audio-processing, communications, tomography, etc. of reconstructing a discrete signal from its samples. The importance of discrete signals has increased dramatically over the years, since most signal processing is done on digital computers, using the discrete representation of the signal. In his paper, Landau ([25]) proves that a necessary condition for reconstruction an irregularly sampled continuous signal is the following:

Let $X = \{f(x_j), j = 1,...,N\}$ be a set of samples of a bandlimited function $f(t) \in L^2(\mathbb{R}^d)$ with finite support $\hat{f} \subseteq \Omega \subseteq \mathbb{R}^d$, where $\hat{f}$ denotes the Fourier transform of $f$. If $f(t)$ can be reconstructed from $X$, then the Beurling density:

$$D(X) = \liminf_{r \to \infty} \inf_{x \in R^d} \frac{\text{card}(X \cap B_r(x))}{(2r)^d} > |\Omega| \qquad \textbf{(4-1)}$$

Where $B_r(x_0) = \{x \in \mathbb{R}^d : |x - x_0|_\infty \leq r\}$ is a **cube** (because of the maximum norm) with edge length $2r$ and volume $|B_r(x_0)| = (2r)^d$.

Equation (4-1) is a necessary condition on the density which says that the sampling density must be higher than the Nyquist density. For the 1D case, this necessary condition is also sufficient. Another proof for Landau's theorem can be found in ([26]).

In [27] A. Aldourbi and K. Grochenig developed a theory for sampling and reconstructing signals from non-uniformly sampled data in shift invariant spaces. In this paper, it is shown that any set of samples in a shift invariant space satisfy the frame condition, and therefore is a set of sampling. Though, the innovation is an extension of the continuous sampling theory to weighted $L^p$ spaces, the development is limited to the Fourier domain. In this chapter, necessary and sufficient conditions for perfect reconstruction of bandlimited (not necessarily in the Fourier domain) discrete signals will be introduced focusing on the special cases of Fourier and Haar in both 1D and 2D case. This form of signal reconstruction, gives the best approximation in the RMS sense for any given transform domain. The transform domain is an a-priori information, in which the reconstruction takes place on. For instance, CT images should be reconstructed on Radon domain, and some images should be reconstructed on the Haar domain (for some images Haar transform has better energy compaction).

## 4.1 Theoretical background for the 1D case

Any 1D band-limited discrete signal can be written as following:

$$a_k = \sum_{r \in BW} \alpha_r \varphi_{r,k} \tag{4-2}$$

Where $a_k, \{k=1,..,N\}$ represents a value of the signal with finite length N, at a certain point k, $\alpha_r$ are the coefficients of the basis functions $\varphi_{r,k}$ within a certain band $BW$. Equation (1) can be written in the following matrix form:

$$\underline{a} = \mathbf{\Phi} \underline{\alpha} \tag{4-3}$$

Where $\underline{a}$ is a column vector of length $N_0$ whose elements are an arbitrary subset of $\{a_k\}_{k=1}^{N}$ (the samples of the signal). The matrix $\mathbf{\Phi}$ is an $N_0 \times \text{card}(BW)$ matrix, where "card" stands for *cardinality*, i.e. the number of elements in a set. Each row has elements $\varphi_{r,k}$ where k is constant along the row, and $r \in BW$.

If card(BW)=$N_0$ then $\mathbf{\Phi}$ is a $N_0 \times N_0$ square matrix. A necessary and sufficient condition for finding the coefficients of the basis functions $\{\alpha_r\}$ and reconstructing the signal is that $\mathbf{\Phi}$ is a non-singular matrix.

*Lemma 1*: Let $N_0 = \text{card(BW)}$ be the number of samples of a band limited signal $a$, and $\mathbf{\Phi}$ is a $N_0 \times N_0$ square matrix as described in equation (2). The signal can be perfectly reconstructed if and only if $\det|\mathbf{\Phi}| \neq 0$

*Lemma 2*: Let $\kappa = \text{rank}(\mathbf{\Phi})$ be the *effective number of samples*. Then a band limited signal can be perfectly reconstructed if and only if $\kappa = \text{card(BW)}$. If $N_0 = \kappa$ then $\mathbf{\Phi}$ is square with $\det|\mathbf{\Phi}| \neq 0$ and the matrix is non-singular. In that case, the coefficients can be found simply by inverting the matrix and solving equation (2). If $N_0 > \kappa$ then excessive rows should be removed from the matrix $\mathbf{\Phi}$ to obtain a new matrix $\tilde{\mathbf{\Phi}}$ so that $\text{rank}(\tilde{\mathbf{\Phi}}) = \text{card(BW)} = N_0$ and $\det|\tilde{\mathbf{\Phi}}| \neq 0$.

If $\kappa < \text{card(BW)}$ then there are not enough samples for the reconstruction of the signal. However, the closest approximation within BW will be obtained from given samples.

The error of the reconstruction, by choosing $N_0$ spectral coefficients, is given by the Parseval relation for orthonormal transforms. The band-limited signal $\hat{\mathbf{A}}_N$ approximates complete signal $\mathbf{A}_N$ with mean squared error:

$$MSE = \|\mathbf{A_N} - \hat{\mathbf{A}}_\mathbf{N}\| = \sum_{k=0}^{N-1} |a_k - \hat{a}_k|^2 = \sum_{r \notin \mathbf{BW}} |\alpha_r|^2 \qquad (4\text{-}4)$$

The MSE error can be minimized by selecting a set of basis functions with the best energy compaction for the signal.

For testing and analyzing the theory, a computer program was written in order to synthesize band-limited signals, sampling them randomly and reconstructing by solving the linear equations. The program also tells the rank of the obtained matrix, so it is possible to know the effective number of samples.

The pseudo code of the program is as follows:

1. Generate a vector (or a matrix, in the 2D case) **s** of white noise; in this case it was i.i.d random variables with uniform distribution.
2. Apply orthonormal transformation **R** to the white noise, if the noise is 2D the operator **R** should be apply to its rows and coloums, so the data after this stage is **Rs** in the 1D case, or $\mathbf{RsR}^T$ in the 2D case.
3. Zero high frequency/scale coefficients and apply inverse transform to obtain a band-limited signal.
4. Sample the signal randomly.
5. Build the linear system as in equation $\underline{a} = \mathbf{\Phi}\underline{\alpha}$

    (4-3).
6. Calculate the rank. If the rank of the matrix is equal to the number of non-zero coefficients (or bigger, in the case of over-sampling), the signal can be reconstructed perfectly by solving the system of equations.
7. Solve the system of equations and restore the complete signal.

## 4.2 Analysis for Fourier basis functions in 1D.

In this section a necessary and sufficient condition for sampling and reconstructing in the Fourier domain will be introduced. In this special case, the basis functions are given by:

$$\varphi_{r,k} = \exp(\frac{j2\pi rk}{N}) \tag{4-5}$$

Where $N$ is the length of the signal. Let $a_k$ be a one-dimensional band-limited signal. Let's assume that BW={-M,...,M} and $N_0 = \text{card(BW)} = 2M+1$ samples of $a_k$ are given. The necessary condition for having enough samples exists here, but this is not sufficient, since the matrix must also be non-singular. Let K be the set of $N_0$ indices of the samples, so that $K = \{k_1,...,k_{N_0}\}, \ k_i \in \mathbb{N}, \ 1 \le k_i \le N\}$ and the given samples are $\{a_{k_1},...,a_{k_{N_0}}\}$

Substituting in equation (2) gives:

$$\begin{pmatrix} a_{k_1} \\ a_{k_2} \\ ... \\ a_{k_{N_0}} \end{pmatrix} = \begin{pmatrix} \exp\left(\frac{j2\pi k_1 *(-M)}{N}\right) & \exp\left(\frac{j2\pi k_1 *(-M+1)}{N}\right) & ... & \exp\left(\frac{j2\pi k_1 * M}{N}\right) \\ \exp\left(\frac{j2\pi k_2 *(-M)}{N}\right) & \exp\left(\frac{j2\pi k_2 *(-M+1)}{N}\right) & ... & \exp\left(\frac{j2\pi k_2 * M}{N}\right) \\ ... & ... & ... & ... \\ \exp\left(\frac{j2\pi k_{N_0} *(-M)}{N}\right) & \exp\left(\frac{j2\pi k_{N_0} *(-M+1)}{N}\right) & ... & \exp\left(\frac{j2\pi k_{N_0} * M}{N}\right) \end{pmatrix} \begin{pmatrix} \alpha_{-M} \\ \alpha_{-M+1} \\ ... \\ \alpha_M \end{pmatrix}$$

(4-6)

Showing that $\boldsymbol{\Phi}$ is a nonsingular matrix, can be found by calculating its determinant in the following way:

$$\det \boldsymbol{\Phi} = \exp\left(\frac{-j2\pi M \sum_{i=1}^{N_0} k_i}{N}\right) \det \begin{pmatrix} 1 & \exp\left(\frac{j2\pi k_1}{N}\right) & ... & \exp\left(\frac{j2\pi k_1 * 2M}{N}\right) \\ 1 & \exp\left(\frac{j2\pi k_2}{N}\right) & ... & \exp\left(\frac{j2\pi k_2 * 2M}{N}\right) \\ ... & ... & ... & ... \\ 1 & \exp\left(\frac{j2\pi k_{N_0}}{N}\right) & ... & \exp\left(\frac{j2\pi k_{N_0} * 2M}{N}\right) \end{pmatrix}$$

(4-7)

The matrix on the right side is a Vandermonde with $\alpha_i = \exp\left(\frac{j2\pi k_i}{N}\right)$ matrix. Its determinant is known and given by $\prod_{1 \le m < n \le N_0} \left(\exp\left(\frac{j2\pi k_m}{N}\right) - \exp\left(\frac{j2\pi k_n}{N}\right)\right)$, so that $\det \boldsymbol{\Phi} \ne 0$ and rank($\boldsymbol{\Phi}$)=$N_0$ for every set of sampling $K$.

**Conclusion:** On Fourier domain, there is no importance for the *location* of the samples; the only demand is to have enough samples. For the special case of a band-limited *real* signal, with one-side bandwidth M, it is required to have 2M+1 samples which can take place anywhere (the "+1" is because of the zero frequency). Similar conclusion can be derived identically, when the signal is bandlimited in the sense of high frequency components.

Once the coefficients are found, the whole signal $a_k, \forall k$ can be found by direct calculation of equation (1).

Figure 4-1 demonstrates sampling and reconstruction on Fourier domain, using the computer program mentioned above.

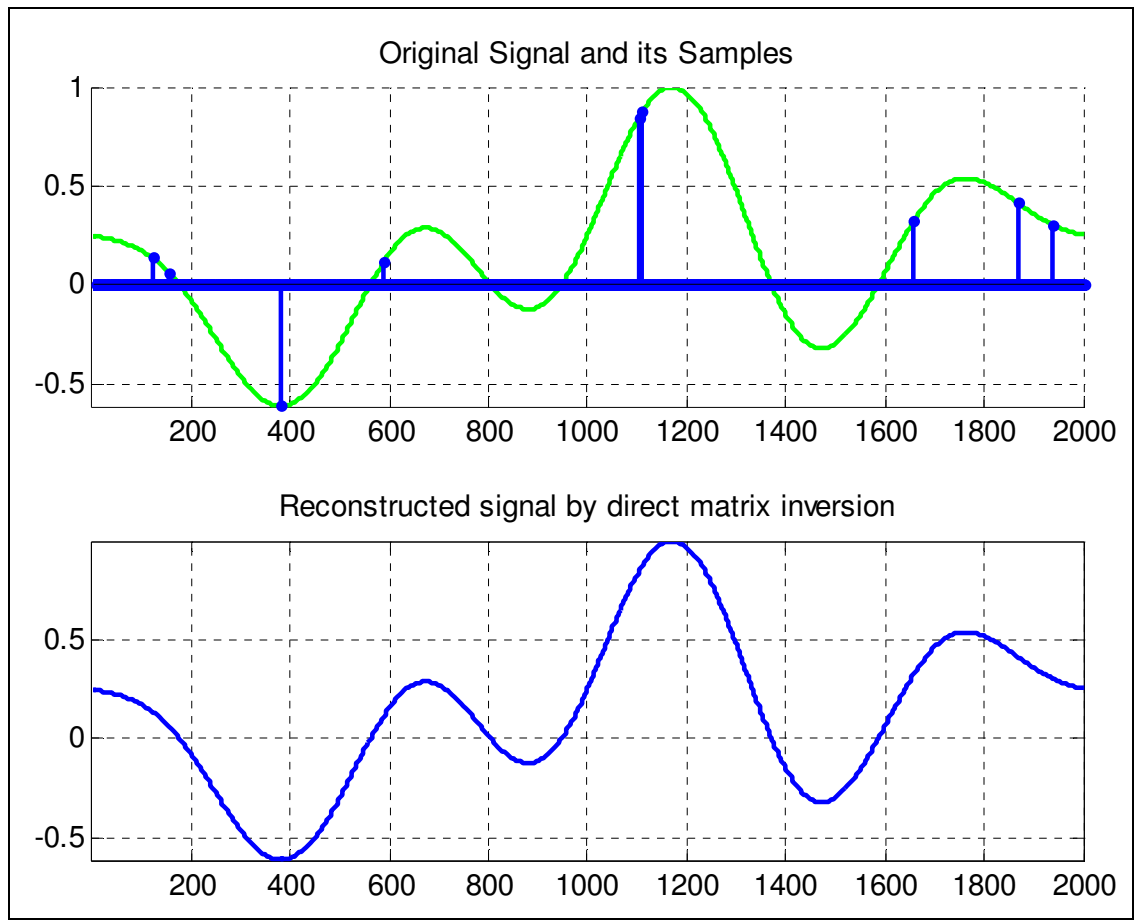

**Figure 4-1: Sampling and reconstructing in Fourier domain**

Figure 4-1 shows a discrete signal of length 2000 sampled randomly at 9 points. In that case $N_0 = \text{card}(BW) = 9$ and since this example is taken on Fourier domain, that automatically implies that $\kappa = 9$ and therefore perfect reconstruction can be achieved.

As for the Discrete Cosine Transform (DCT), since N-point DCT of a signal is equivalent to 2N-point Shifted Discrete Fourier Transform (SDFT) with shift parameters (1/2,0) of 2N- sample signal obtained from the initial one by its mirror reflection ([28]). The $\Phi$ matrix of SDFT(1/2,0) is given by:

$$\Phi_{DCT} = \begin{pmatrix} \exp\left(\frac{j2\pi(k_1 + 1/2)*(-M)}{2N}\right) & \exp\left(\frac{j2\pi(k_1 + 1/2)*(-M+1)}{2N}\right) & \cdots & \exp\left(\frac{j2\pi(k_1 + 1/2)*M}{2N}\right) \\ \exp\left(\frac{j2\pi(k_2 + 1/2)*(-M)}{2N}\right) & \exp\left(\frac{j2\pi(k_2 + 1/2)*(-M+1)}{2N}\right) & \cdots & \exp\left(\frac{j2\pi(k_2 + 1/2)*M}{2N}\right) \\ \cdots & \cdots & \cdots & \cdots \\ \exp\left(\frac{j2\pi(k_{N_0} + 1/2)*(-M)}{2N}\right) & \exp\left(\frac{j2\pi(k_{N_0} + 1/2)*(-M+1)}{2N}\right) & \cdots & \exp\left(\frac{j2\pi(k_{N_0} + 1/2)*M}{2N}\right) \end{pmatrix}$$

4-8)

Using the fact that $\Phi_{DCT} = \Phi_{DFT-2N} * diag\left\{\exp\left(j\pi \frac{M_i}{2N}\right)\right\}$ and applying same logic as above to show that $\Phi_{DFT-2N}$ is always invertible, leads to the conclusion that $\Phi_{DCT}$ is always invertible as well.

### 4.3 Analysis for Haar wavelet basis functions in 1D

Wavelets are basis functions that are scaled and shifted. The scale is analog to frequency in each scale has several shifts in it. If $N_0 = 2^k$, then the signal will have $k$ scales. Wavelets are sometimes suggested when working with non-stationary signals (for instance Linear FM or *"chirp"* signal) or if the signal has discontinuities. Haar functions are based on rectangular functions, and they're a special case of the Daubechies wavelet with 2 coefficients of the father wavelet, sometimes normalized to $\left(\frac{1}{\sqrt{2}}, \frac{1}{\sqrt{2}}\right)$ in order to have a sum of $\sqrt{2}$ and a square sum of 1. Haar wavelet is also known as D2 wavelet. The basis functions appear below, each basis function is a row in the transform matrix.

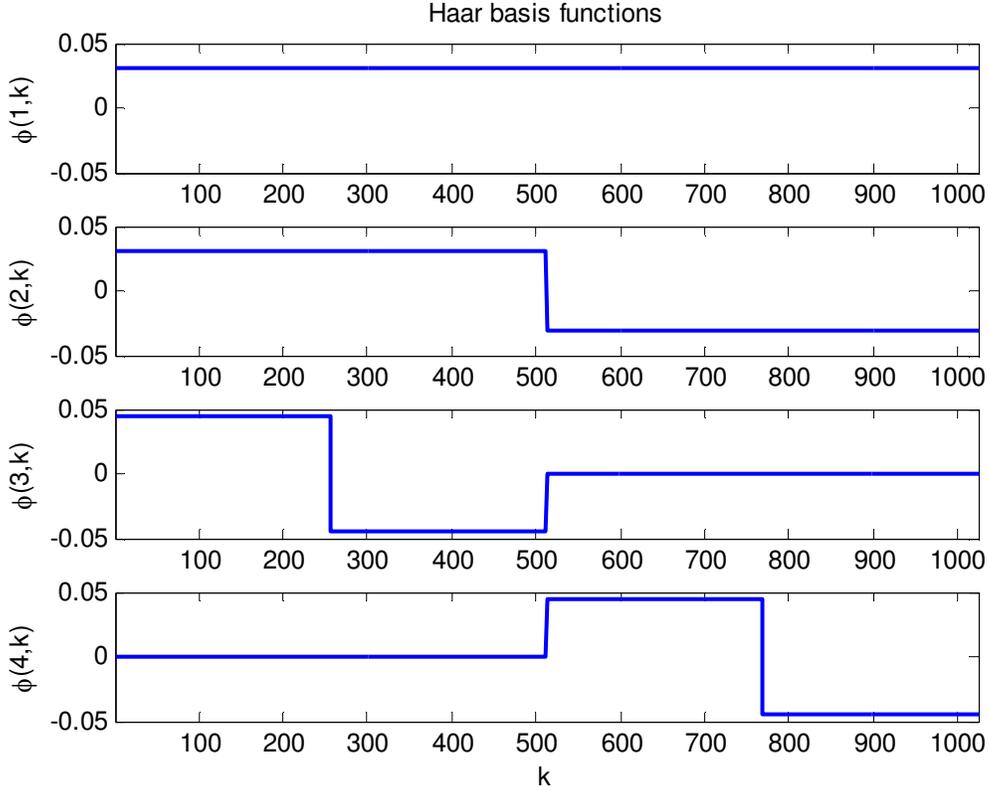

**Figure 4-2: Haar basis functions**

Figure 4-2 shows the basis functions $\varphi_{r,k}$ for different scales and shifts. The first function (the constant) is the *scale*, and the other functions are the *wavelets*. As was shown before, a bandlimited signal on Haar domain can be written as the following:

$$a_k = \sum_{r \in \text{BW}} \alpha_r \varphi_{r,k} \qquad (4\text{-}9)$$

It can be easily seen, from the Haar basis functions, that for some sampling locations k, the basis functions have the same value. For instance, $\varphi_{1,k} = \varphi_{2,k}$ for $k < \dfrac{N}{2}$. Those cases derive identical rows in the matrix $\mathbf{\Phi}$ and therefore $|\mathbf{\Phi}| = 0$ which means that perfect reconstruction cannot be achieved. In other words, unlike in the Fourier case, where the location of the samples wasn't important to achieve perfect reconstruction, on the Haar domain, there is a special condition for reconstruction, since there are sets of sampling locations $K = \{k_1,...,k_{N_0}\}$, $k_i \in \mathbb{N}$, $1 \leq k_i \leq N\}$ so that $\det|\Phi(K)| = 0$.

Let us now examine the special case where $N_0 = \text{card}(BW) = 2$, $K = \{k_1, k_2\}$ on the Haar domain. Id est, the entire signal has 2 spectral coefficients and 2 samples are given and the length of signal is N (N – even). The system of linear equations to be solved is:

$$\begin{pmatrix} a_{k_1} \\ a_{k_2} \end{pmatrix} = \begin{pmatrix} \varphi_{1,k_1} & \varphi_{2,k_1} \\ \varphi_{1,k_2} & \varphi_{1,k_2} \end{pmatrix} \begin{pmatrix} \alpha_1 \\ \alpha_2 \end{pmatrix}$$

(4-10)

In order to solve the equation, it is required that $\det|\Phi(k_1, k_2)| \neq 0$. In other words $\det|\Phi(k_1,k_2)| \neq 0$ if and only if $\left(k_1 \leq \frac{N}{2} \cap k_2 > \frac{N}{2}\right) \cup \left(k_2 \leq \frac{N}{2} \cap k_1 > \frac{N}{2}\right)$, one sample on each half of the signal.

The equation of prohibited sampling sets $\det|\Phi(K)| = 0$ for $N_0 = \text{card}(BW)$ is an $N_0$ dimensional geometric region. In the case above, where $N_0 = 2$ the geometric region that describes the forbidden sampling sets is:

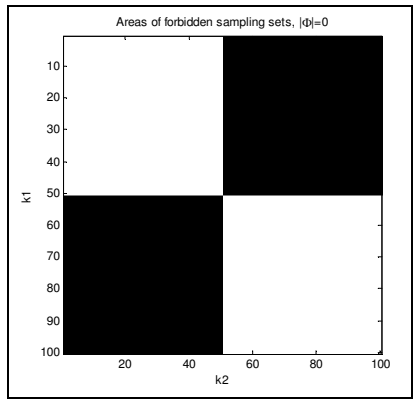

**Figure 4-3: The geometric region in which perfect reconstruction is possible from 2 samples**

Figure 4-3 shows for which sampling locations $(k_1, k_2)$ perfect reconstruction can be achieved. White areas stand for perfect reconstruction, and black areas stand for sets that are inappropriate for perfect reconstruction. In that Haar case, it is required to have at least one sample on every resolution interval. Figure 4-4 illustrates a recoverable and not-recoverable cases on the Haar domain.

The conclusion for 2 samples, can be elegantly be extended for the case where $N_0 = \text{card}(BW) = 2^m, m$ - positive integer. In this special case, the necessary and sufficient condition for perfect reconstruction is to uniformly divide the length N interval of signal length into $N_0$ sub-intervals and each sub-interval must have one sample. The location of the sample within its sub-interval is not important.

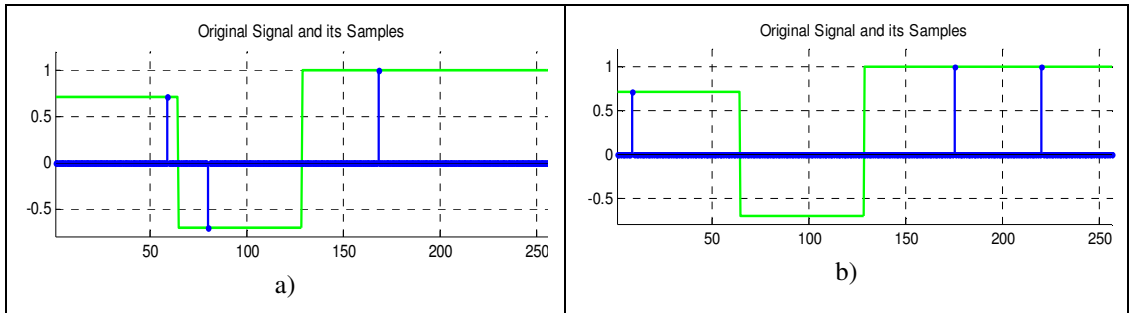

**Figure 4-4: a) Recoverable sampling in the Haar domain b) Not Recoverable sampling on the Haar domain**

Numerical example: $N = 512; N_0 = \text{card}(BW) = 4$, where the transform domain is Haar. Since $N_0 = 4$ the number of scales is 2 (logarithm of 4), with 2 shifts in each scale.

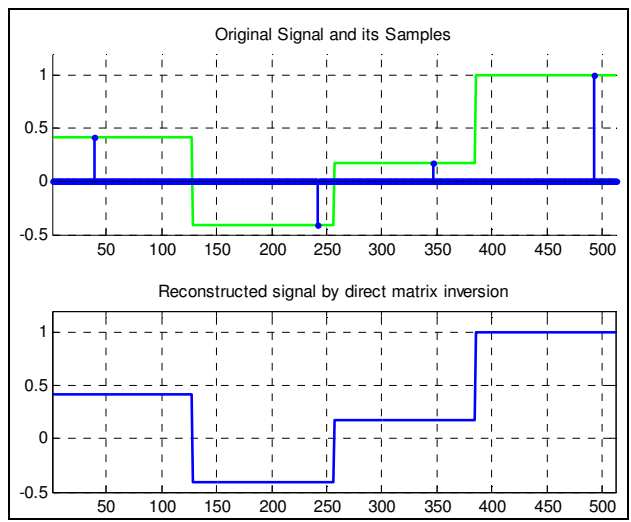

**Figure 4-5: Perfect reconstruction on Haar domain**

Here, the matrix $\boldsymbol{\Phi}$ is:

$$\boldsymbol{\Phi} = \begin{pmatrix} 0.0442 & 0.0442 & 0.0625 & 0 \\ 0.0442 & -0.0442 & 0 & -0.0625 \\ 0.0442 & -0.0442 & 0 & 0.0625 \\ 0.0442 & 0.0442 & -0.0625 & 0 \end{pmatrix} \quad (4\text{-}11)$$

And $\operatorname{rank}(\boldsymbol{\Phi}) = 4$ which means that it is possible to reconstruct the signal perfectly from its samples, as indeed shown in Figure 4-5.

Another Numerical Example: Again, $N = 512; N_0 = \operatorname{card}(\mathrm{BW}) = 4$, but this time the signal cannot be reconstructed, as shown in Figure 4-6.

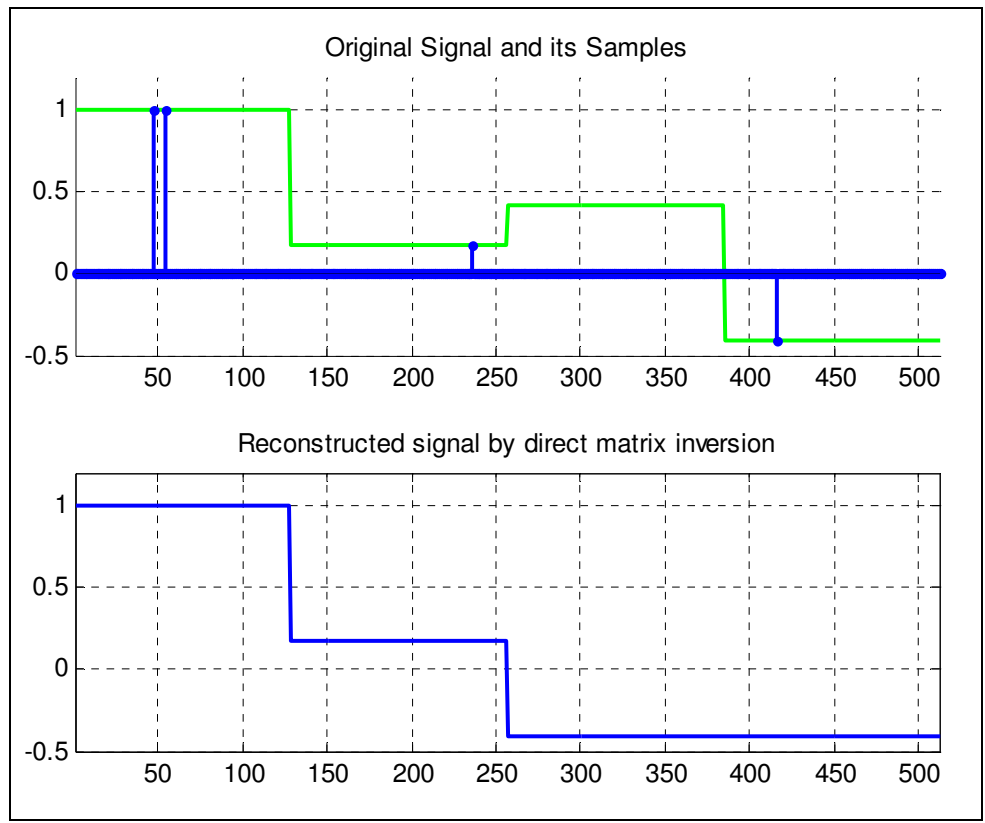

**Figure 4-6: Futile reconstruction using matrix inversion, due to samples'**

In this case:

$$\mathbf{\Phi} = \begin{pmatrix} 0.0442 & -0.0442 & 0 & -0.0625 \\ 0.0442 & 0.0442 & -0.0625 & 0 \\ 0.0442 & 0.0442 & 0.0625 & 0 \\ 0.0442 & 0.0442 & 0.0625 & 0 \end{pmatrix}$$ (4-12)

And $\text{rank}(\mathbf{\Phi}) = 3$, which means that the *effective number* of samples in this example was 3, where it is required to have 4, and therefore perfect reconstruction cannot be achieved. Instead, the best band-limited approximation in the RMS sense was obtained.

It can be seen in the first example where perfect reconstruction was feasible, that there was a sample in each interval of length 512/4=128 samples and in the second example the third interval [257,384] wasn't sampled.

## 4.4 Analysis for the 1D, D4-wavelet basis functions

Another example from the Daubechies wavelet family, is the D4 wavelet. Like Haar, this wavelet is also orthogonal with compact support, but this time has 4 father wavelet coefficients, which are the following (normalized to have a sum of $\sqrt{2}$ ):

$$h_n = \begin{pmatrix} 0.4829629131445341 \\ 0.8365163037378077 \\ 0.2241438680420134 \\ -0.1294095225512603 \end{pmatrix}$$

More about the reconstruction of the Daubechies wavelets can be found in ([29]).
Since the basis functions are not piece-wise constant, the reconstruction condition is more relaxed compared to the Haar wavelet case, in which usually two close samples have the same values in their basis functions which lead to a two identical rows in the matrix, and therefore made it non-invertible.

Numerical Example: Signal length: 512, $N_0 = 16$ (log16=4 scales), and 16 samples. In this example, reconstruction was perfect as can be shown in Figure 4-7.

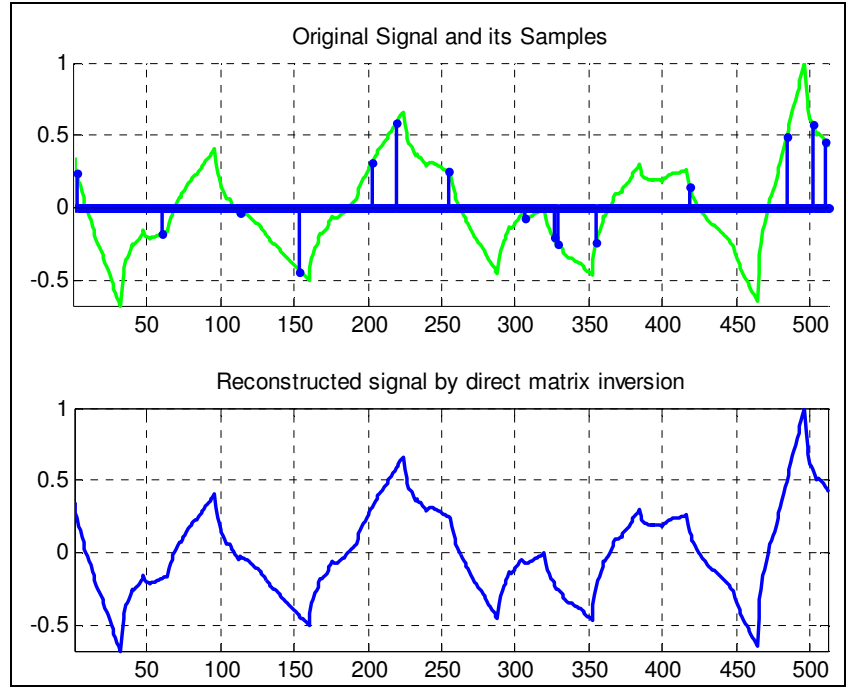

**Figure 4-7: Perfect reconstructed on the D4-wavelet domain**

## 4.5 Analysis for 1D Walsh basis functions

Another set of interesting and useful basis functions, are the Walsh basis functions. Walsh transform has been widely used in digital signal processing, image processing, communications (CDMA and Spread Spectrum) and logic design. Before going any farther, it is required to define the Walsh matrix, in which each row is a Walsh function.

The Hadamard matrices of dimension $2^k$, $k \in \mathbb{N}$ are given by the recursive formula:

$$\begin{aligned}
\mathbf{H}_1 &= 1 \\
\mathbf{H}_2 &= \begin{bmatrix} 1 & 1 \\ 1 & -1 \end{bmatrix} \\
\mathbf{H}_{2^k} &= \begin{bmatrix} \mathbf{H}_{2^{k-1}} & \mathbf{H}_{2^{k-1}} \\ \mathbf{H}_{2^{k-1}} & -\mathbf{H}_{2^{k-1}} \end{bmatrix} = \mathbf{H}_2 \otimes \mathbf{H}_{2^{k-1}}
\end{aligned}$$

(4-13)

Where $\otimes$ denotes the Kronecker product.

The sequence ordering of the rows of the Walsh matrix is obtained by applying bit reversal permutation and then Gray code permutation. Bit reversal is the permutation where the data at an index *n* (starting from zero), written in binary with digits $b_4b_3b_2b_1b_0$ (e.g. 5 digits for *N*=32 inputs), is transferred to the index with reversed digits $b_0b_1b_2b_3b_4$. Gray code permutation is similar, except that the index is transferred to the binary value of the index written in Gray code. (Transferring from binary to Gray is done with the formula: G = B XOR (SHR(B)), where SHR is shift right once, and XOR is the Exclusive-OR logical operation). The Walsh basis functions are similar to Haar functions. Yet, since Walsh is not a wavelet, it does not have shifts and it gets only values of 1 and -1, where Haar can also have value of zero. Note that the Walsh functions are similar to binary Fourier functions of sines and cosines with symmetry (even) or anti-symmetry (odd) with respect to the middle of the signal. The index of the basis functions corresponds to "sequency", or to the number of zero crossing of the basis function.

Figure 4-8 shows the first 8 Walsh basis functions.

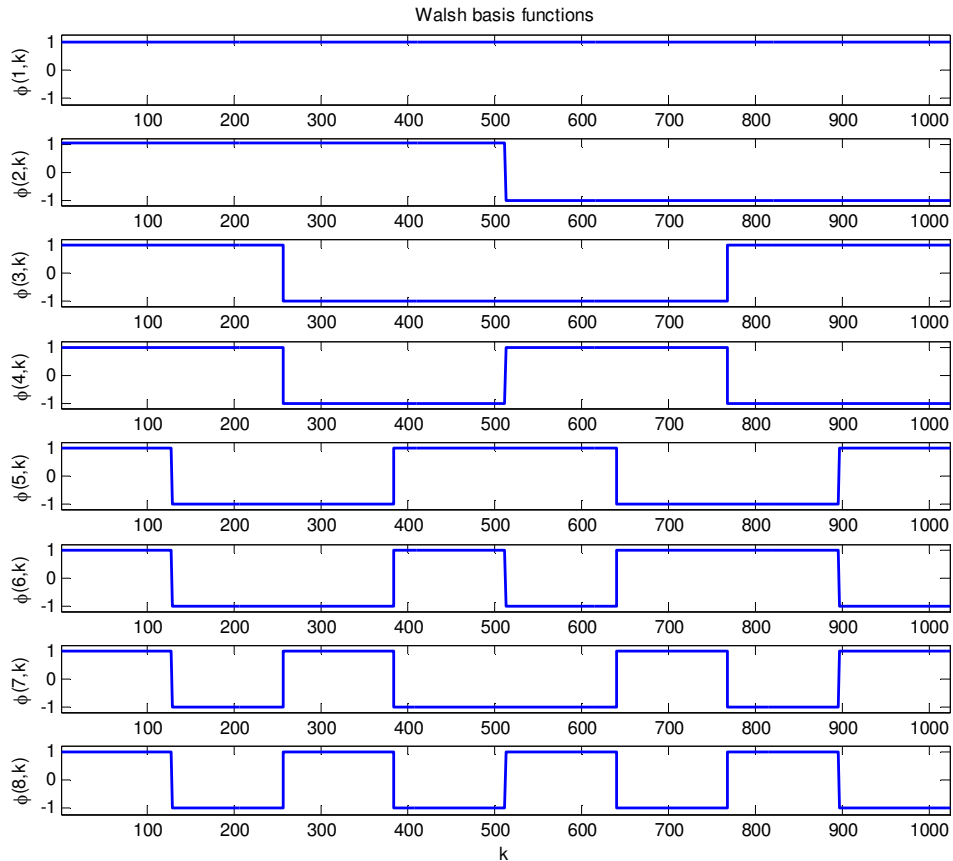

**Figure 4-8: The first 8 Walsh basis functions**

Again, each basis functions is a row in the transform matrix, except that the transform matrix should be normalized by $\frac{1}{\sqrt{n}}$ so that $\mathbf{HH}^T = \mathbf{I}$. It can be seen from the figure, that the <u>sufficient</u> condition for perfect reconstruction is similar to the one of Haar, except that this time it is not necessary. The reason that the condition for Walsh is less strict compared with Haar, is because that if Haar is not sampled in each of its intervals (shifts), then the matrix will have a column of zeros, where in the Walsh case, there are no zeros in the basis functions so the matrix is not invertible only when its rows (columns) are linear dependent. For instance, assuming a signal is composed of 3 Walsh basis functions then the matrix will have 3 rows. From the basis functions, we know that each row is one of the rows in the following matrix:

$$\begin{pmatrix} 1 & 1 & 1 \\ 1 & 1 & -1 \\ 1 & -1 & -1 \\ 1 & -1 & 1 \end{pmatrix} \qquad (4\text{-}14)$$

Each row represents the values of the 3 basis functions on a different sampling and equal sampling intervals ([1,..,N/4], [N/4+1,…,N/2], etc.) Obviously any 3 out of 4 of these (row) vectors are linearly independent, and therefore as long as the 3 samples are taken from 3 different intervals (out of 4) it is possible to achieve perfect reconstruction. Note, that this means that the interpolation in that case is not nearest neighbor. In general, for any $N_0$, the number of different value basis functions intervals is $2^{\lceil \log_2 N_0 \rceil}$, so for $N_0 = 3,4$ there are 4 intervals, for $N_0 = 5,6,7,8$ there are 8, and so on. In order to have perfect reconstruction, it is necessary (and not always sufficient) to have the $N_0$ samples taken from different intervals. So if the signal length is $N = 1024$ and $N_0 = 5$, then there are 8 intervals, each one is 128 samples long. For perfect reconstruction, the intervals that contain the 5 samples must be distinct. Note that for the special case of $N_0$ the reconstruction condition is to have a sample in each interval, just like in the Haar case.

Numerical example: $N = 512, N_0 = \text{card(BW)} = 5$

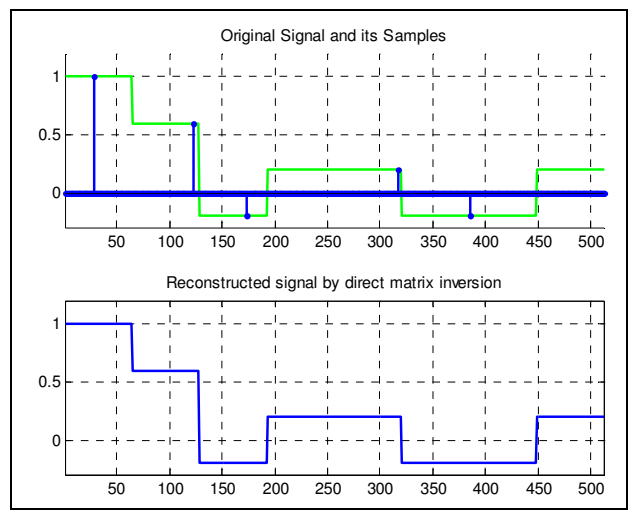

**Figure 4-9: Example for perfect reconstruction on Walsh domain**

In this example obtained matrix was:

$$\Phi = \begin{pmatrix} 0.0442 & -0.0442 & 0.0442 & -0.0442 & -0.0442 \\ 0.0442 & -0.0442 & -0.0442 & 0.0442 & 0.0442 \\ 0.0442 & 0.0442 & 0.0442 & 0.0442 & 0.0442 \\ 0.0442 & 0.0442 & -0.0442 & -0.0442 & -0.0442 \\ 0.0442 & 0.0442 & 0.0442 & 0.0442 & -0.0442 \end{pmatrix} \quad \text{(4-15)}$$

And its rank equals 5 and indeed perfect reconstruction was obtained with zero error. It is interesting to note, that though the Walsh basis functions are similar to Haar functions, if the signal in this example was bandlimited on Haar domain, the perfect reconstruction couldn't occur, since one of the intervals has no samples at all.

## 4.6 Theoretical background for the 2D case

The theory developed for 1D signals, can be easily extended to the multidimensional case, except that now the basis functions are n-dimensional. More specifically, a 2D signal can be written as following:

$$a_{k_1,k_2} = \sum_{r_1 \in BW_1} \sum_{r_2 \in BW_2} \alpha_{r_1,r_2} \varphi_{r_1,r_2,k_1,k_2} \qquad (4\text{-}16)$$

Or in a matrix form as:

$$\underline{a} = \Phi \underline{\alpha} \qquad (4\text{-}17)$$

Assuming $\text{card}(BW_1) = N_0$, $\text{card}(BW_2) = M_0$ and the number of samples is $Q$ then $\Phi$ is a $Q \times Q$ matrix, which means that for perfect reconstruction, a necessary condition is $Q \geq N_0 M_0$. Once the matrix is found, the conditions for perfect reconstruction remain the same as in the 1D case. Id est, the number of effective samples is $\kappa = \text{rank}(\Phi)$ and the condition on samples can be described by the equation $\det|\Phi(K)| = 0$ where K is the sampling set of the signal, where in this case each sampling has 2 co-ordinates, since now the signal is 2D, $K = \{(m_1, n_1), ..., (m_Q, n_Q)\}$ (Q samples). If $\kappa = N_0 M_0$ then there are enough samples for perfect reconstruction. In the analysis brought here, the assumption is that signal band-limitation is separable (rectangular spectrum).

## 4.7 Analysis for Fourier basis in the 2D case

Assuming our 2 dimensional signal (image) has size of $N \times N$ then its basis functions on Fourier domain are $\varphi_{r_1, r_2, k_1, k_2} = \exp(j\frac{2\pi}{N} r_1 k_1 + j\frac{2\pi}{N} r_2 k_2)$ then the image can be expressed as:

$$\underline{a} = \sum_{r_1 \in BW_1} \sum_{r_2 \in BW_2} \alpha_{r_1, r_2} \varphi_{r_1, r_2, k_1, k_2} = \sum_{r_1 \in BW_1} \sum_{r_2 \in BW_2} \exp(j\frac{2\pi}{N} r_1 k_1 + j\frac{2\pi}{N} r_2 k_2); \ k_1, k_2 = 1, ..., N \quad \text{(4-18)}$$

The above equation can be written in the compact matrix form:

$$\underline{a} = \mathbf{\Phi} \underline{\alpha} \quad \text{(4-19)}$$

Assuming the image is sampled so only Q samples are given, then $\mathbf{\Phi}$ is a $Q \times Q$ matrix. Now, let the image $\underline{a}$ be band-limited on both axes, so that $\text{card}(BW_1) = N_0$, $\text{card}(BW_2) = N_0$ and Q is the minimum number of samples, then $Q = N_0^2$ and $\mathbf{\Phi}$ is a $N_0^2 \times N_0^2$ matrix. This means, that for relatively narrow band images, the matrix $\mathbf{\Phi}$ can be quite large.

In the 2D Fourier case of $Q = N_0^2$ and $\text{card}(BW_1) = N_0$, $\text{card}(BW_2) = N_0$ the matrix $\mathbf{\Phi}$ is invertible for every sampling set, so like in the 1D case, the location of the samples is not important.

<u>Numerical example:</u> Here the image size was 64x64, and $\text{card}(BW_1) = \text{card}(BW_2) = 23$ so 529 samples are needed for perfect reconstruction, the blue dots denote the sampling points. The original test image and its samples is given in Figure 4-10(a)

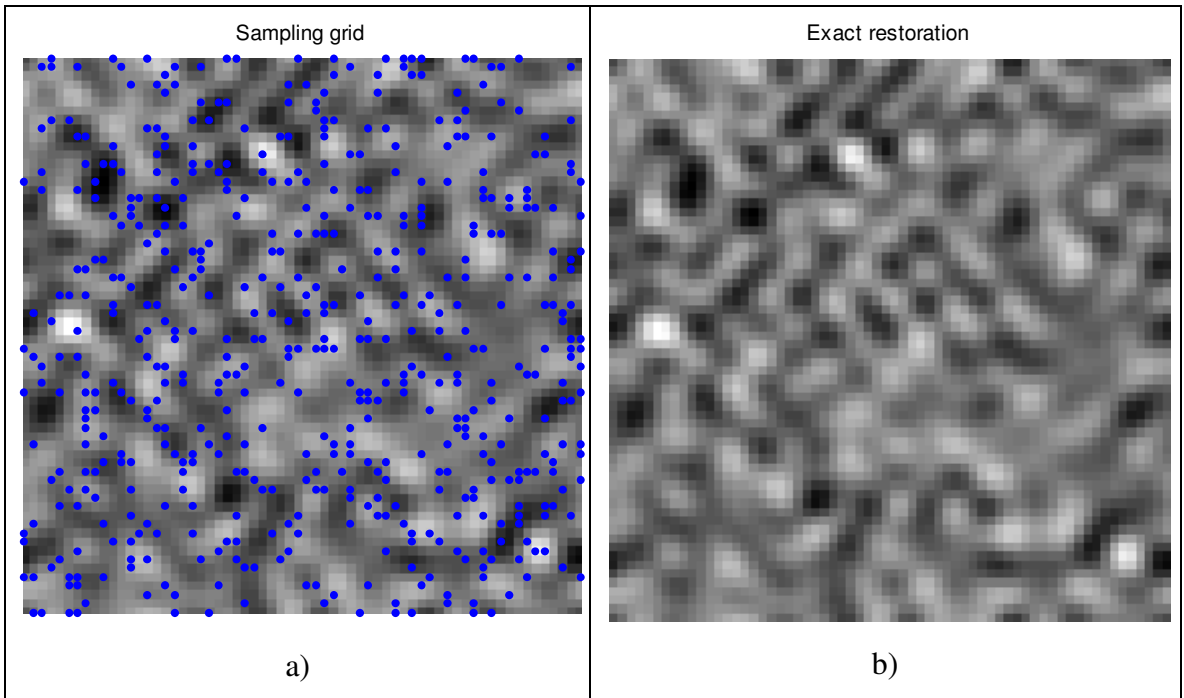

**Figure 4-10: a) Band-limited (Fourier domain) random image sampled randomly (samples are given by the blue dots), b) Reconstructed image from the given samples**

The matrix size was 529x529. By inverting the matrix the spectral coefficients can be found and then the entire image can be reconstructed, as shown in and Figure 4-10(b). The reconstruction is exact, with error zero.

## 4.8 Analysis for the 2D Haar basis:

In the 2D case, the Haar basis functions are again shifted and scaled and it is a separable function of the 1D Haar, so that $\varphi_{x,y}^{2D} = \varphi_x \varphi_y^T$, the transform is given by $\mathbf{Y} = \mathbf{HXH}^T$

An example for several 2D Haar basis functions can be seen below:

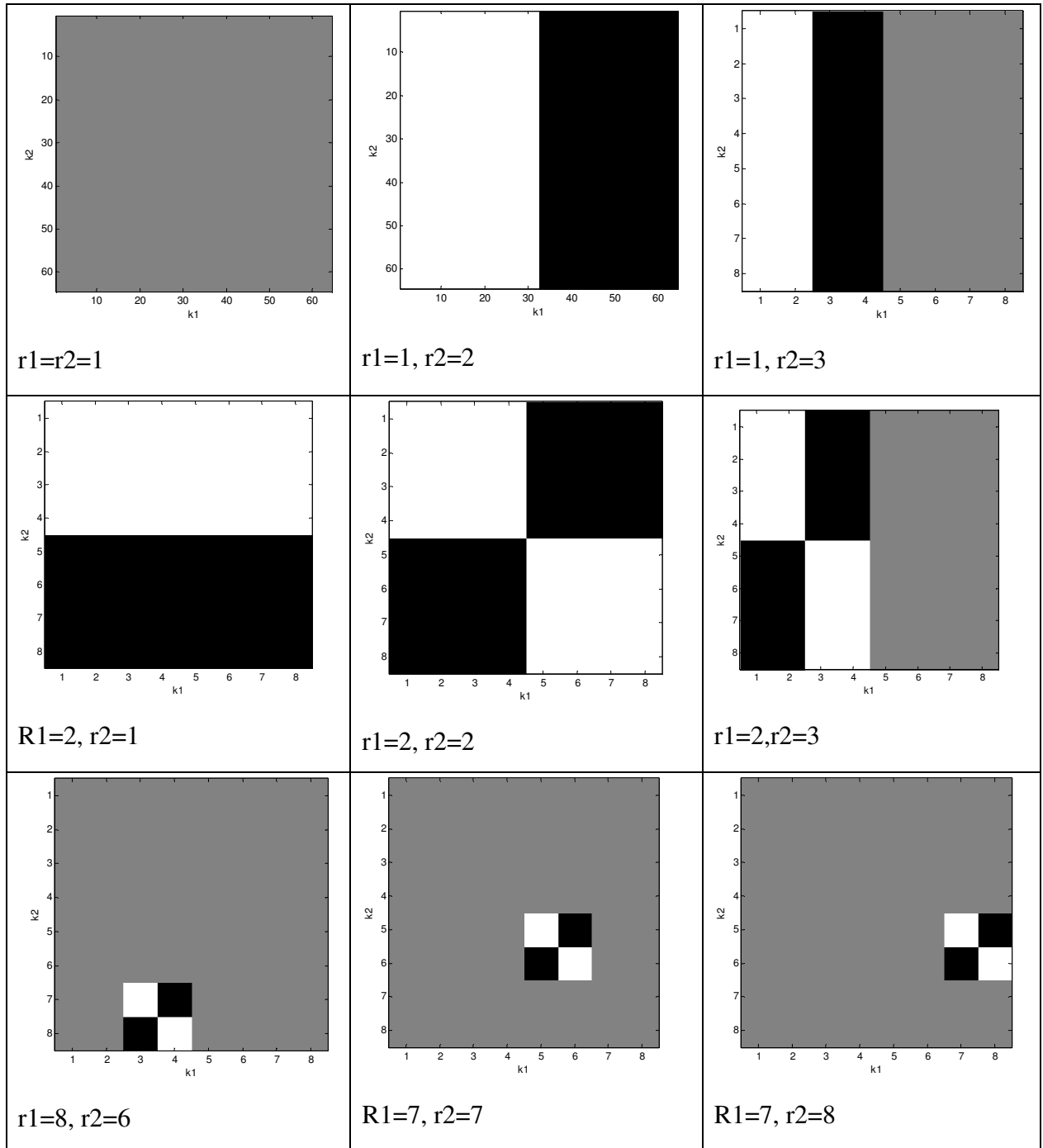

**Figure 4-11: Some 2D Haar basis functions**

In the case shown above, 8 1D Haar basis functions were taken, to produce 64 2D Haar functions, 9 were shown in Figure 4-11. Since the 2D Haar functions are separable, it is clear that the same condition on sampling from the 1D case also applied to the 2D case. If the in 1D case there were intervals that different basis functions had the same value, then the same occurs in the 2D case: there are rectangles that get the same value for different functions as can be seen from the functions above, or directly from the fact that the Haar function is separable. A necessary and sufficient condition for sampling and reconstructing on the Haar basis, is to have a sample on every rectangle of the Haar function.

Numerical example: In this example, a band limited on Haar domain image was created. N=64 (for every axis, 4096 pixels all in all), N0=8 (again, for each axis) and 64 samples were taken randomly, marked by blue dots. In this case shown in Figure 4-12(a), it is easy to see that the sampling weren't taken in an appropriate position. For perfect reconstruction one sample must be found on every one of 8x8 fragments that generate the whole image. It can be seen from the figure, that there are fragments which haven't been sampled (for instance the upper left, the three squares on the upper right and some others). In that case, the rank of the matrix was 41 (affective number of samples) instead of 64 required for perfect reconstruction. However, in the case shown in Figure 4-12(b), there's a sample in every square, and the signal is recoverable. In the Haar case, the interpolation is trivial – nearest neighbor.

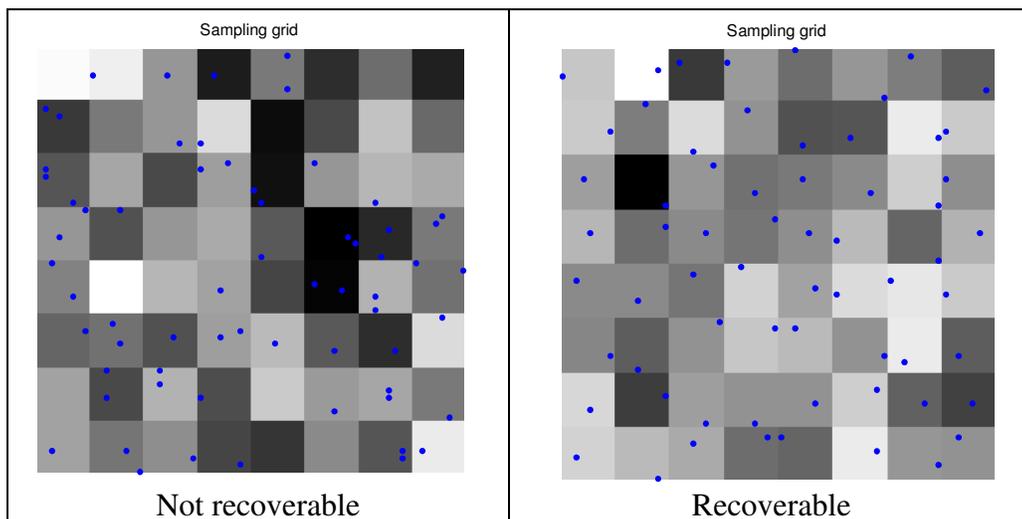

**Figure 4-12: Band limited (Haar domain) sampled test images. a) Not recoverable, b) Recoverable**

## 4.9 Analysis for the 2D Walsh basis:

Similarly to the Haar and Fourier basis, the Walsh basis functions can be extended to the 2D case simply by separability: $\varphi_{x,y}^{2D} = \varphi_x \varphi_y^T$. Several 2D Walsh basis functions are presented in Figure 4-13. Walsh basis functions are similar to Haar functions, except that they are binary (1, -1) and are not a wavelet. If $N_0$ is a power of two, then the reconstruction condition is similar to the Haar case.

*2D Walsh functions:*

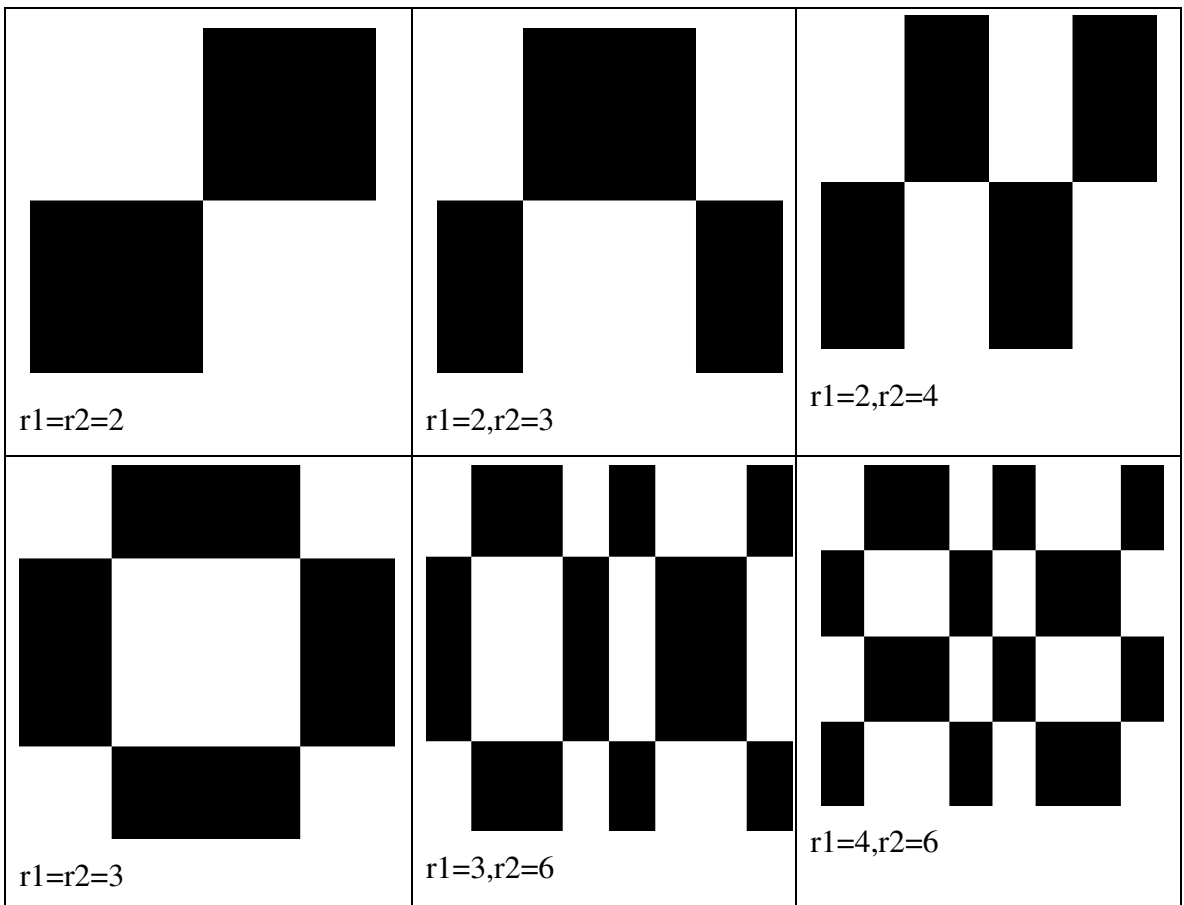

**Figure 4-13: Some 2D Walsh basis functions**

In the example brought in Figure 4-14 a band limited on Walsh domain image was created. N=64, N0=5 and 25 samples were taken. On one scenario the image could be restored perfectly, and on the other it couldn't, because of the location of the samples.

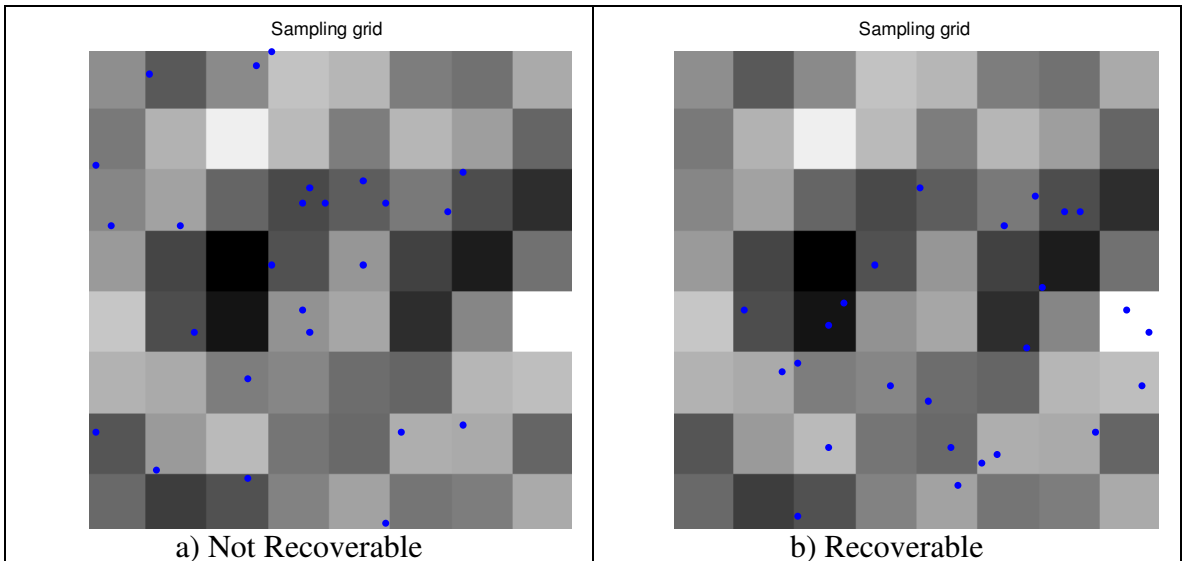

**Figure 4-14: Two cases of sparse sampling of an image band-limited in Walsh Transform domain: a) not recoverable case; b) recoverable case (sample points are marked with dots). Image size was 64x64, and band-limitation was 5x5.**

Just like in the 1D case, the image was reconstructed using the Walsh basis functions. The same image could also be represented using the Haar basis functions, but obviously in this case it would require more than 25 coefficients, because some of the squares are not sampled, and Haar reconstruction is simply nearest neighbor interpolation.

## 4.10 Iterative Reconstruction

In real life applications, signal reconstruction (or best approximation, when perfect reconstruction is impossible, as was discussed above) by means of matrix inversion is often not practical since usually the obtained matrix is huge and inverting it is impossible due to computer resources or due to the time required. Instead, numerical methods, such as Gauss-Siedel or conjugate gradients for solving the linear set of equation can be used. The Papoulis-Gerchberg interpolation ([23]) that was used in the SR algorithm is an iterative process for estimating the Fourier transform of a band-limited function if only a segment (or segments) of the entire function is given. The reconstruction is done by applying the band-limiting operator $\mathbf{B} = \mathbf{F}^{-1}\mathbf{V}\mathbf{F}$, where $\mathbf{F}$ is the Fourier transform operator and $\mathbf{V}$ is a sampling operator other than $\mathbf{I}$ (usually an ideal low-pass filter where its band-pass is from an a-priori knowledge)**.** After the operator is applied to the given function, the known samples from the function are replaced in the obtained function, and the process is repeated (i.e. applying again the operator **B** and replacing the known segments again and so on). This process can be seen as a procedure that solves the linear set of equations and gives the best approximation in terms of $l_2$-norm inside a given bandwidth (or $L_2$-norm for the continuous case). In [24] a similar process is suggested for interpolation of signals and it is shown that the restriction for perfect reconstruction is only a cardinality restriction: for any given bandwidth, there's a maximum percentage of lost samples that can be tolerated. Obviously, as long as the bandwidth increases, the required percentage required for perfect reconstruction also increases. Moreover, the Papoulis-Gerchberg interpolation can be applied not only in the Fourier domain, but to other invertible transform domains as well, such as orthogonal wavelets and Radon. This makes it a powerful tool for perfect reconstruction (if possible) and signal approximation which of course depends on the number of iterations taken for the process.

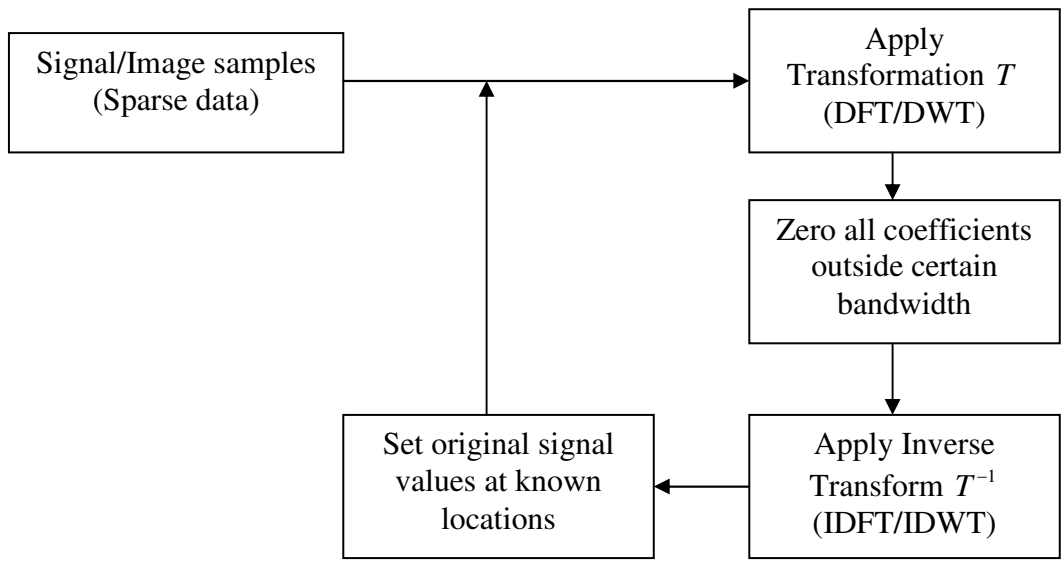

**Figure 4-15: Block diagram of the iterative reconstruction algorithm**

Numerical example:

In this case, a 256 samples discrete signal with $\text{card}(BW) = 5$ was sampled 5 times and iterative interpolation was applied and 10,000 iterations were used. The results after taking 52 and after taking 10,000 iterations are shown in Figure 4-16. Having more samples than needed, will cause the iterative process to converge faster. Also, the location of the samples influences the speed of convergence. Having more uniformly distributed samples will converge faster, than samples concentrated on a single area.

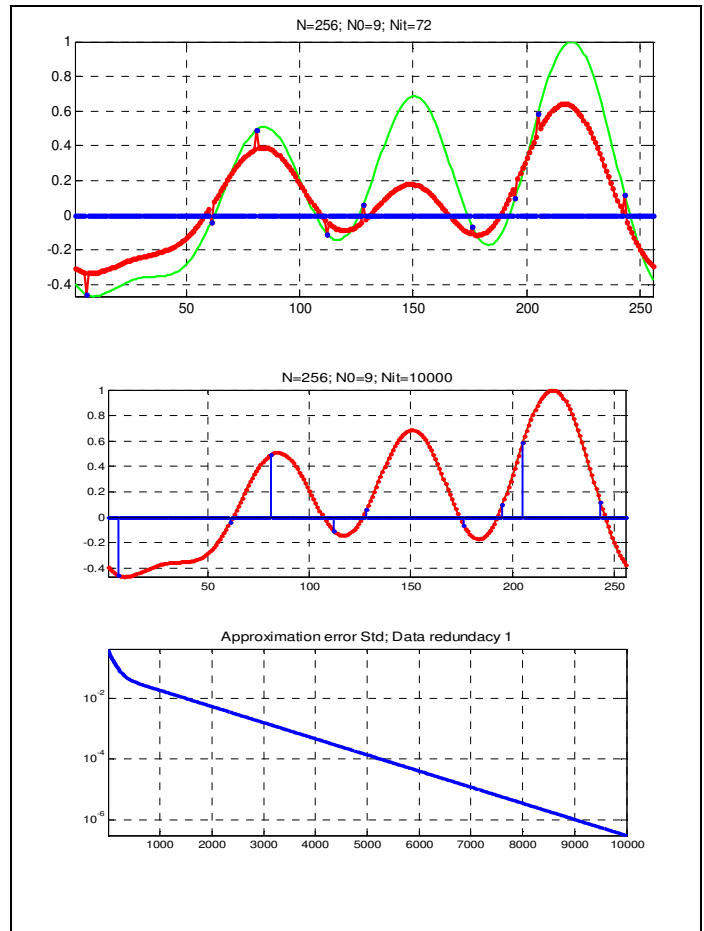

**Figure 4-16: Reconstruction on Fourier domain using the iterative algorithm**

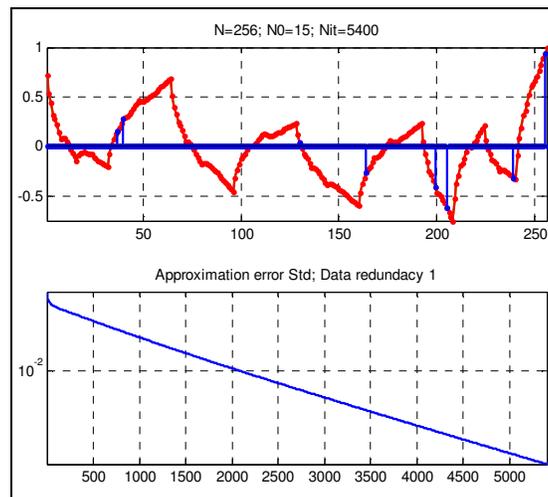

**Figure 4-17: iterative reconstruction of the D4-wavelet domain**

The error at the end of the process is less than $10^{-6}$ which indicates the feasibility of perfect reconstruction. Figure 4-17 illustrates the same process on the D4-wavelet domain.

As was mentioned before, the location of the samples affects the speed of convergence. If the samples are close to one another, concentrated in a small area, then the algorithm will take more time to converge. Figure 4-18 shows the results of a comparison between two cases for the same signal (length: 64, 7 samples): in the first one, illustrated in Figure 4-18a, the samples are grouped together and the convergence is very slow, after 500 iterations the error is larger than $10^{-4}$, where in the second case, were the samples are more uniformly distributed, illustrated in Figure 4-18b, the error is smaller than $10^{-10}$ 10^-10! This comparison illustrates the major importance the location of the samples has. Since the transform domain is Fourier, then in both cases, exact reconstruction could be achieved by direct matrix inversion.

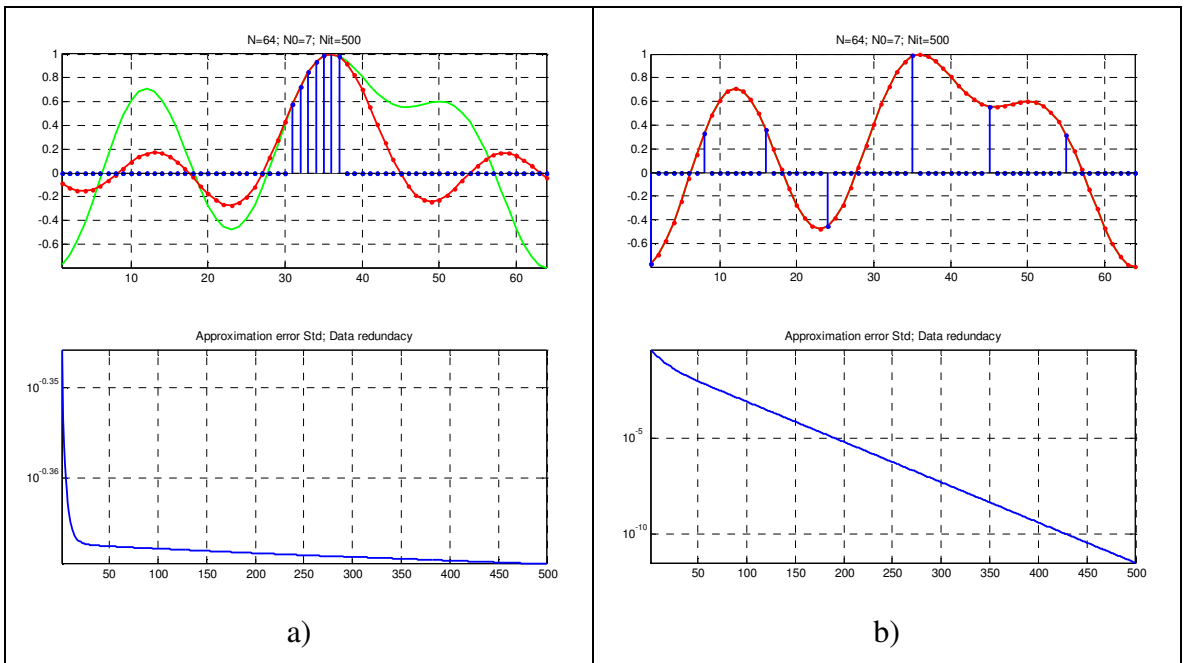

**Figure 4-18: Comparison between the convergence of two types of sampling a) grouped samples b) uniformly distributed samples**

An applicative example for the theory is recovery of an image corrupted by "salt & pepper" noise. In case of the "salt & papper" noise, image pixels are replaced, with a certain probability of error, with their extreme values that correspond to signal minimum or maximum. With certain probability of false alarms, erroneous pixels can be quite easily detected, and the distorted image can be subjected to the above described iterative reconstruction procedure that will generate a band-limited approximation of the image that preserves available not distorted pixels, the band limitation being determined by the rate of non-distorted pixels. Figure 4-19a) and b) illustrate an example of such restoration for the case when the probability of a missing pixel is 0.5. The resulting image shown in Figure 4-19b) is an approximation to the original image that is low pass band-limited in DCT domain by a 90° circle sector of radius 0.7 (in the units of the base band).

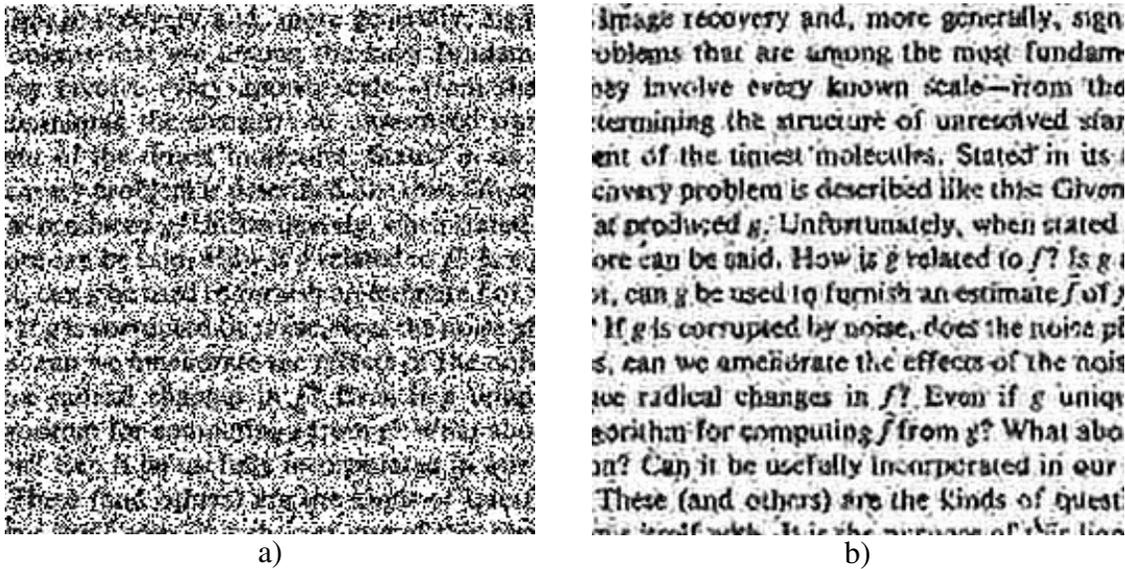

a) b)

**Figure 4-19: Recovery of images corrupted by "salt & pepper" noise with probability of pixel missing equal to 0.5.**

# 5 IMAGE RECONSTRUCTION OF COMPUTER TOMOGRAPHY

## 5.1 The Radon transform

The discrete sampling theory from chapter 4 can be used to improve medical Computer Tomography (CT) images. Computed Tomography is based on the x-ray principal: as x-rays pass through the body, they are absorbed or attenuated (weakened) at different levels creating a matrix or profile of x-ray beams of different strength. This x-ray profile called a sinogram which forms the raw data that can be converted to an image of an internal organ.

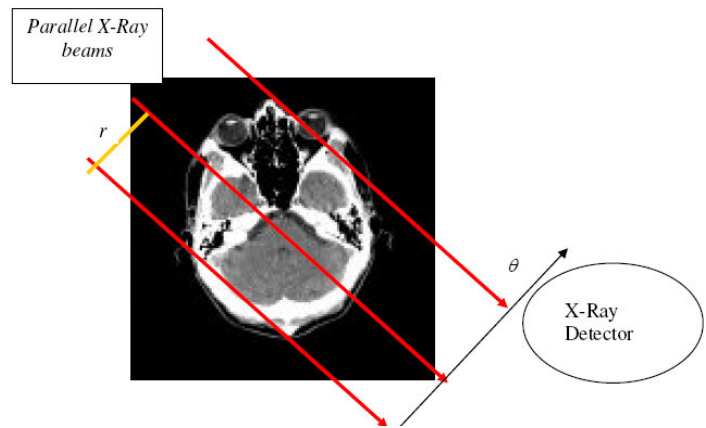

**Figure 5-1: Radon Transform**

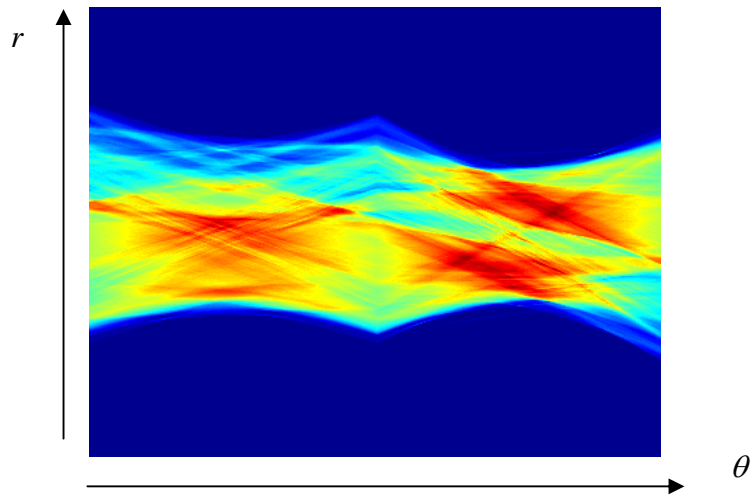

**Figure 5-2: The sinogram of the image from Figure 5-1**

The mathematical description of the process that creates parallel projections from an image is the continuous Radon transform:

$$RI(r,\theta) = \int_{-\infty}^{\infty} \int_{-\infty}^{\infty} I(x,y)\delta(x\cos\theta + y\sin\theta - r)dxdy \qquad (5\text{-}1)$$

Where $I(x,y)$ represents the image and $RI(r,\theta)$ represents its sinogram (its Radon transform). The inverse Radon transform can be found by using the projection theorem.

Applying 1D Fourier transform to the above equation gives the projection theorem:

$$\int_{-\infty}^{\infty} RI(r,\theta)\exp(-2\pi i f_\theta r)dr = \int_{-\infty}^{\infty}\int_{-\infty}^{\infty} I(x_1, x_2)\exp\left[2\pi i\left(f_\theta x_1 \cos\theta + f_\theta x_2 \sin\theta\right)\right]dx_1 dx_2 \qquad (5\text{-}2)$$

The expression on the right hand side of the equation is actually 2D Continuous Polar Fourier transform. Changing variables and applying 2D inverse Fourier transform gives:

$$I(x,y) = \int_0^\pi \int_{-\infty}^{\infty} |f_\theta| RI(f_\theta,\theta)\exp\left[-2\pi i f_\theta (x\cos\theta + y\sin\theta)\right] df_\theta d\theta \qquad (5\text{-}3)$$

As can be seen, it is possible to exactly inverse the Radon transform and to create an image from its samples. The problem begins with the discrete case, in which interpolation is needed when going along a diagonal line. This calculation is not exact, and therefore the discrete Radon transform cannot be inverted straight forward. However, iterative method for calculating inverse Radon transform to any desired degree of accuracy has been developed by A. Averbuch *et al.* ([30],[31]) based on the back-projection theorem and polar FFT. The typical accuracy after 3 iterations of the inverse Radon transform is 6 digits. The number of iterations that was taken here was 4, which gives even better accuracy. Since the transform $\mathbf{R}: I \to R$ is one to one, there is a bounded operator $\mathbf{R}^\dagger : R \to I$ so that $\mathbf{R}^\dagger \mathbf{R} = \mathbf{Id}$. Although an algorithm for finding the inverse transform is not given in ([30],[31]), a fast iterative exact algorithm is brought. The algorithm for finding the inverse Radon transform is based on the fact that the Radon transform is a composition of (inverse) Fourier transform and Pseudo-polar Fourier transform (PPFT), $\mathbf{R} = \mathbf{F}_1^{-1}\mathbf{P}$ where $\mathbf{F}_1^{-1}$ is the inverse 1D Fourier transform operator, and $\mathbf{P}$ is the pseudo polar Fourier transform operator. From this, it is clear to see that finding the inverse Radon transform is equal to applying 1D Fourier transform followed by applying inverse pseudo polar Fourier transform ($\mathbf{R}^{-1} = \mathbf{P}^{-1}\mathbf{F}_1$).

**Definition:** *Pseudo-Polar Fourier Transform* $\mathbf{P}$ is a linear transformation for data $I(m,n)$ to a data $\mathbf{P}I(1,k,l) = \hat{I}\left(-\frac{2l}{n}k, k\right)$; $\mathbf{P}I(2,k,l) = \hat{I}\left(k, -\frac{2l}{n}k\right)$ where $\hat{I}(k,l)$ is the 2D FFT of the image *I*. The PPFT is actually resampling of the 2D DFT on a pseudo polar grid as shown in Figure 5-3. When the calculation point of the PPFT requires sampling the 2D FFT in a non-integer point, it can be done efficiently by using the Fractional DFT defined by:

$$\mathbf{F}_N^\alpha \mathbf{x} = \sum_{n=0}^{N-1} x[n] e^{\frac{-2\pi i \alpha k n}{N}}, \quad k = 0,...,N-1, \ \alpha \in \mathbb{R} \tag{5-4}$$

And the resampling operator for the PPFT is given by:

$$\mathbf{G}_{n,k} = \tilde{\mathbf{F}}_m^\alpha \mathbf{F}_1^{-1}, \qquad \alpha = \frac{2k}{n} \tag{5-5}$$

Where $\tilde{\mathbf{F}}_m^\alpha$ is an operator that accepts sequence of length n is, symmetrically pads it by zeros to length $m = 2n+1$, applies the fractional Fourier transform with factor $\alpha$ and returns the $n+1$ central elements.

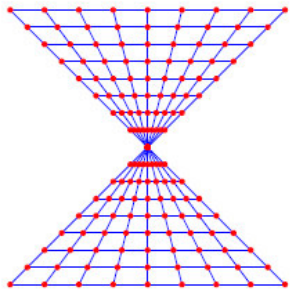 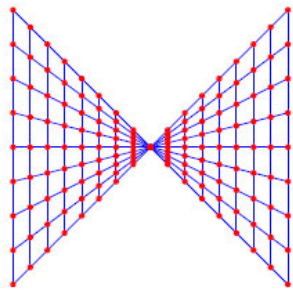 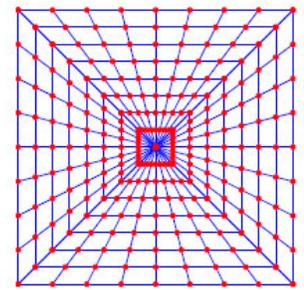

(a) Pseudo-polar sector $\Omega_{pp}^1$
(b) Pseudo-polar sector $\Omega_{pp}^2$
(c) Pseudo-polar grid $\Omega_{pp} = \Omega_{pp}^1 \cup \Omega_{pp}^2$

**Figure 5-3: Pseudo polar grid**

A pseudo-code for an algorithm that calculates the PPFT and its adjoint $\mathbf{P}^*$ can be found in ([32]), where the adjoint can found simply by reversing the algorithm for the PPFT and replacing every operation by its adjoint, which is very simple for an FFT based algorithm.

Now after the PPFT and its link to the Radon transform and its inverse were introduced, an inverse Radon transform can be found by doing the following:

Let $Y = \mathbf{R}I$ be the radon transform of the image I.

1. Define $F = \mathbf{F}_1 Y$, the one dimensional discrete Fourier Transform of Y, as in

$$I(x, y) = \int_0^\pi \int_{-\infty}^\infty |f_\theta| RI(f_\theta, \theta) \exp\left[-2\pi i f_\theta (x\cos\theta + y\sin\theta)\right] df_\theta d\theta \qquad (5\text{-}3)$$

2. Define $\tilde{I} = \mathbf{P}^* F$, back project to the image domain using the adjoint PPFT transform

3. Solve iteratively the Hermitian system: $\mathbf{P}^*\mathbf{P}I = \tilde{I}$

A simple way for understanding the algorithm, is simply by applying $\mathbf{P}^*\mathbf{F}_1$ to both hands of the equation $Y = RI$, giving $\mathbf{P}^*\mathbf{F}_1 Y = \mathbf{P}^*\mathbf{F}_1 \mathbf{R}I = \mathbf{P}^*\mathbf{P}I$ using the fact that $\mathbf{F}_1 \mathbf{R} = \mathbf{P}$

Steps 1-2 are accomplished exactly, whereas step 3 is not exact and not given in a closed form. The operator on the left hand of step 3 is positive, self-adjoint and bounded by its smallest and largest eigenvalues, finding the original image can be done iteratively using the fact that $\mathbf{P}^*\mathbf{P}$ is symmetric and positive definite and therefore $\lambda_{\min}\mathbf{Id} \leq \mathbf{P}^*\mathbf{P} \leq \lambda_{\max}\mathbf{Id}$ where $\lambda_i$ are the eigenvalues of the operator $\mathbf{P}^*\mathbf{P}$, then it can be shown (For instance see: [29], pp. 61-63) that:

$$I = \left(\mathbf{P}^*\mathbf{P}\right)^{-1} \tilde{I} = \frac{2}{\lambda_{\min} + \lambda_{\max}} \sum_{k=0}^\infty \mathbf{R}^k \tilde{I} \qquad (5\text{-}6)$$

Where:

$$\mathbf{R} \triangleq \left( \mathbf{Id} - \frac{2}{\lambda_{min} + \lambda_{max}} \mathbf{P}^* \mathbf{P} \right) \tag{5-7}$$

Using the equations above, the image can be computed iteratively, by the following:

$$I_N = \frac{2}{\lambda_{min} + \lambda_{max}} \tilde{I} + \mathbf{R} I_{N-1} \tag{5-8}$$

The "conjugate gradient" method can also be applied for solving $\mathbf{P}^*\mathbf{P}I = \tilde{I}$. The conjugate gradient method is an algorithm for the numerical solution of particular linear systems. For solving the system: $\mathbf{Ax} = \mathbf{b}$ where $\mathbf{A}$ is symmetric and positive definite (in this case: $\mathbf{A} \triangleq \mathbf{P}^*\mathbf{P}$), the solution is given by the pursuing the following steps:

Initialization:
$r_0 := b - Ax_0$
$p_0 := r_0$
$k := 0$
repeat:
$\quad \alpha_k := \frac{r_k^T r_k}{p_k^T A p_k}$
$\quad x_{k+1} := x_k + \alpha_k p_k$
$\quad r_{k+1} := r_k - \alpha_k A p_k$
$\quad$ if $r_{k+1}$ is "sufficiently small" exit loop
$\quad \beta_k := \frac{r_{k+1}^T r_{k+1}}{r_k^T r_k}$
$\quad p_{k+1} := r_{k+1} + \beta_k p_k$
$\quad k := k + 1$
end repeat

The solution of the system is $\mathbf{x}_{k+1}$. More information on the conjugate gradient method can be found in [33]. To accelerate the process of convergence, it is possible to use a preconditioning real diagonal operator $\mathbf{M}$, and to solve the equation $\mathbf{P}^*\mathbf{MP}I = \mathbf{M}\tilde{I}$ the same way discussed above since $\mathbf{P}^*\mathbf{MP}$ is also symmetric and positive definite.

## 5.2 Reconstruction of CT images from sparse data

4 showed that it is possible to reconstruct perfectly a band-limited discrete signal if the set of sampling meets certain conditions. A necessary condition for reconstruction other than having a band-limited signal is to have enough samples: the number of samples must be at least as the number of coefficients in the transform domain (under special cases, such as working in Fourier domain, this necessary condition is also sufficient, as was shown before). This theory states, that under certain condition, it is possible to reconstruct an image from a finite set of projections in which having more projections will not improve the quality of the image. Though image reconstruction of CT images has been done in the past ([34]), the work done here is based on discrete sampling theorem and exact inverse Radon transform, so perfect reconstruction may be achieved.

The fact that CT images have finite support (i.e. have zero value outside a certain area of the image), can be used as an a-priori information. Moreover, it is possible to consider the sinogram (the Radon transform of the image) as a "Dual image" in which is bandlimited on the *inverse* Radon domain (the image). Therefore, interpolation can be applied for a given set of projections to obtain reconstruction. In other words, the process can be seen as solving the set of linear equations as was described in chapter 4, on the inverse Radon domain. The solution can be found by iterative process such as the Gerchberg-Papoulis interpolation on Radon domain, in a similar way to what has been with Fourier and Haar, except that now the CT image plays the role of the (Radon) spectra and the sinogram the role of the interpolated image.

For investigating the reconstruction error that can be achieved, a simulation program was built, whose block diagram appears in Figure 5-4: Computer model block diagram. The program takes a CT input image, removes some of its projections and tries to reconstruct. Then, root mean square error is calculated.

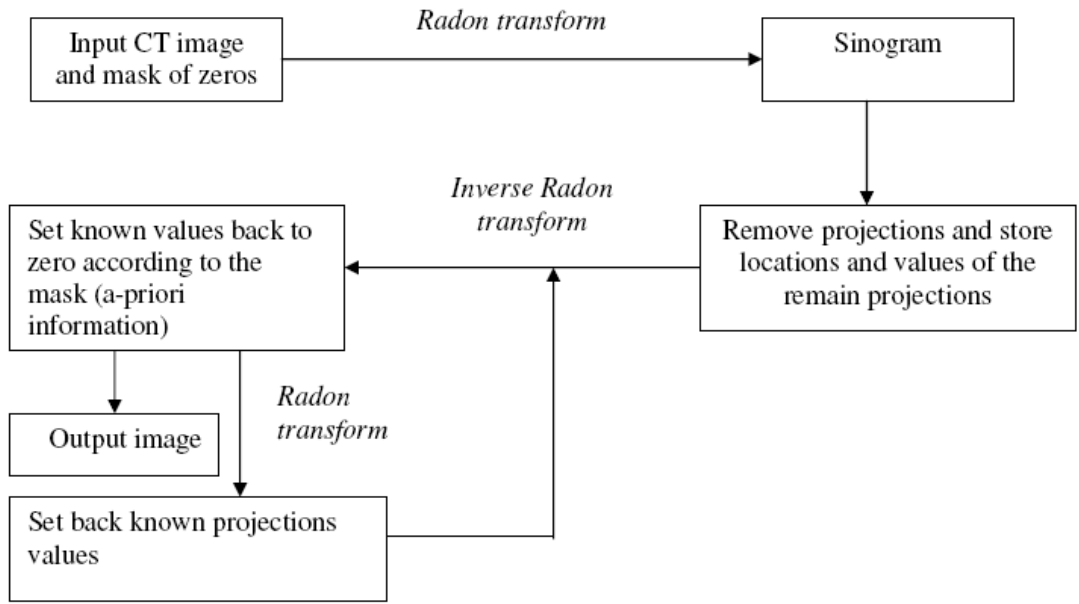

**Figure 5-4: Computer model block diagram**

The CT image from Figure 5-5 was used as a test image for the simulation program for the reconstructing process. 55% of the image is zero, so 55% of the image projections of the Radon transform were randomly removed (set to zero). The image after removing the projections and its Radon transform appear in Figure 5-5(c)-(d).

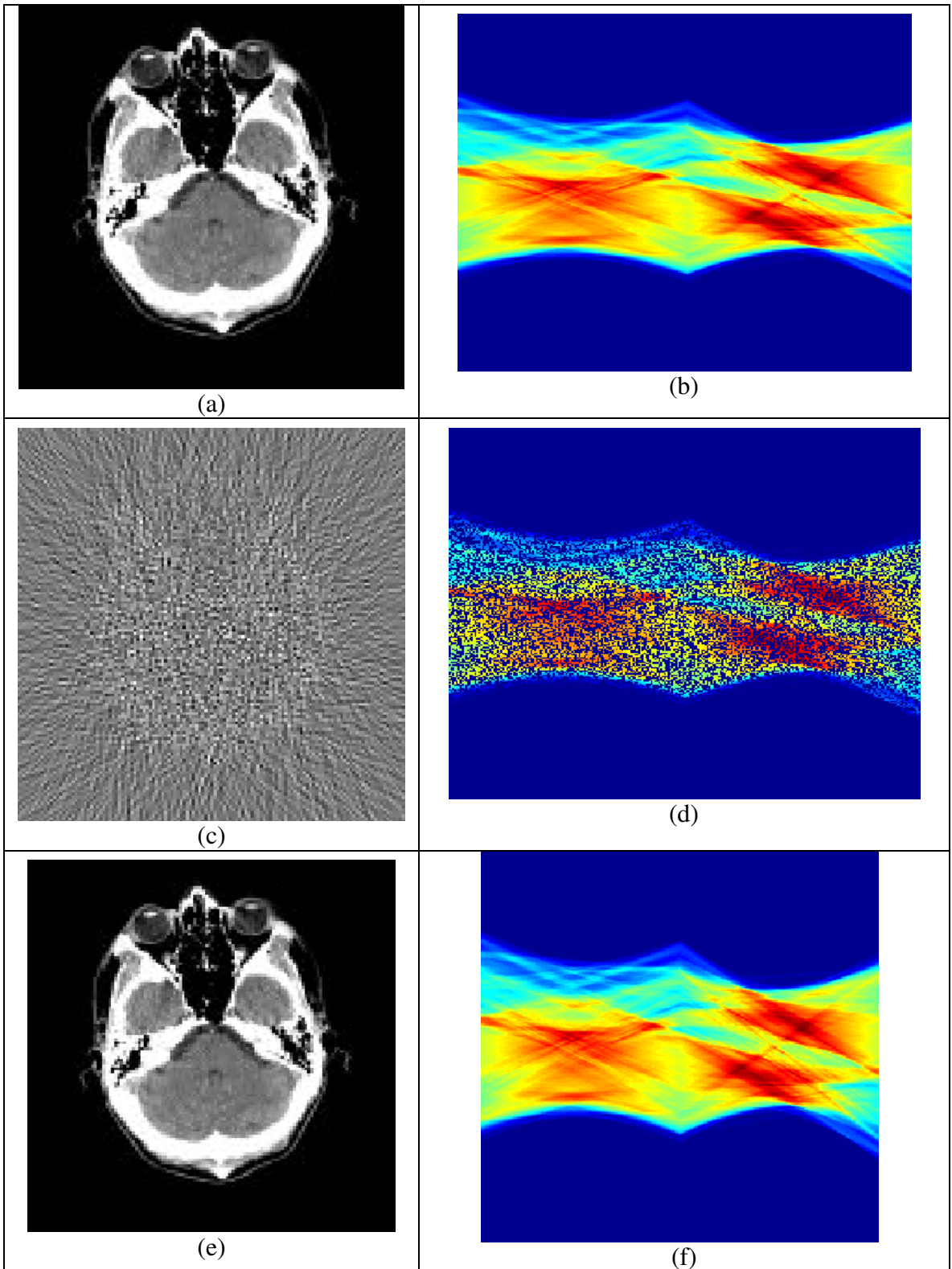

**Figure 5-5: (a)-(b) Original CT image and its Radon transform, (c)-(d) Original CT image after removing 55% of its projections. (e)-(f) Reconstructed CT image**

The value and locations of the known projections is stored and used for the iterative interpolation on Radon domain. The interpolated image is the image appears in Figure 5-5d, which is bandlimited on the inverse Radon domain, as shown in Figure 5-5c. The obtained image and its Radon transform appear in Figure 5-5(e)-(f)

In this example, perfect reconstruction was virtually obtained, as can be seen from figure that shows the error verses the number of iterations. The error in this case is in the order of $10^{-9}$.

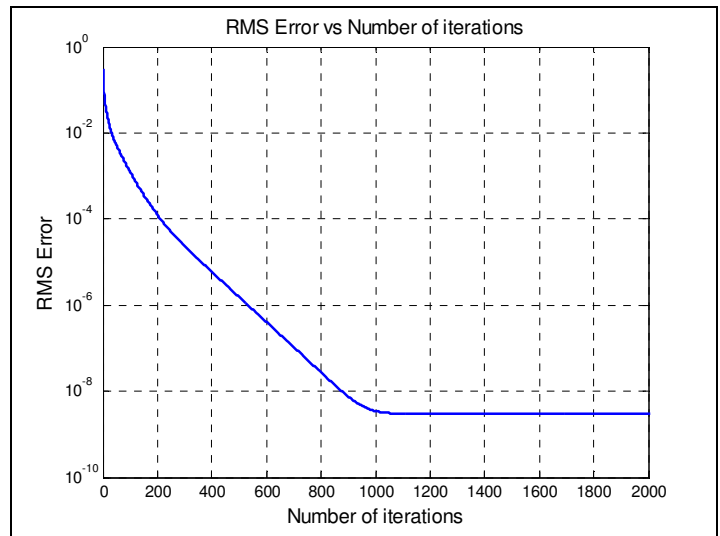

**Figure 5-6: RMS Error**

Another example is given in Figure 5-7. In this example, every second and third projection in the angular axis were removed, so only one third of the entire angles remained. From this set of projections, the original image was reconstructed, based on the black area outside the image as an a-priori image (finite support on the inverse Radon domain) and the value of the real known values from on the Radon domain. The convergence of the process is illustrated in Figure 5-8.

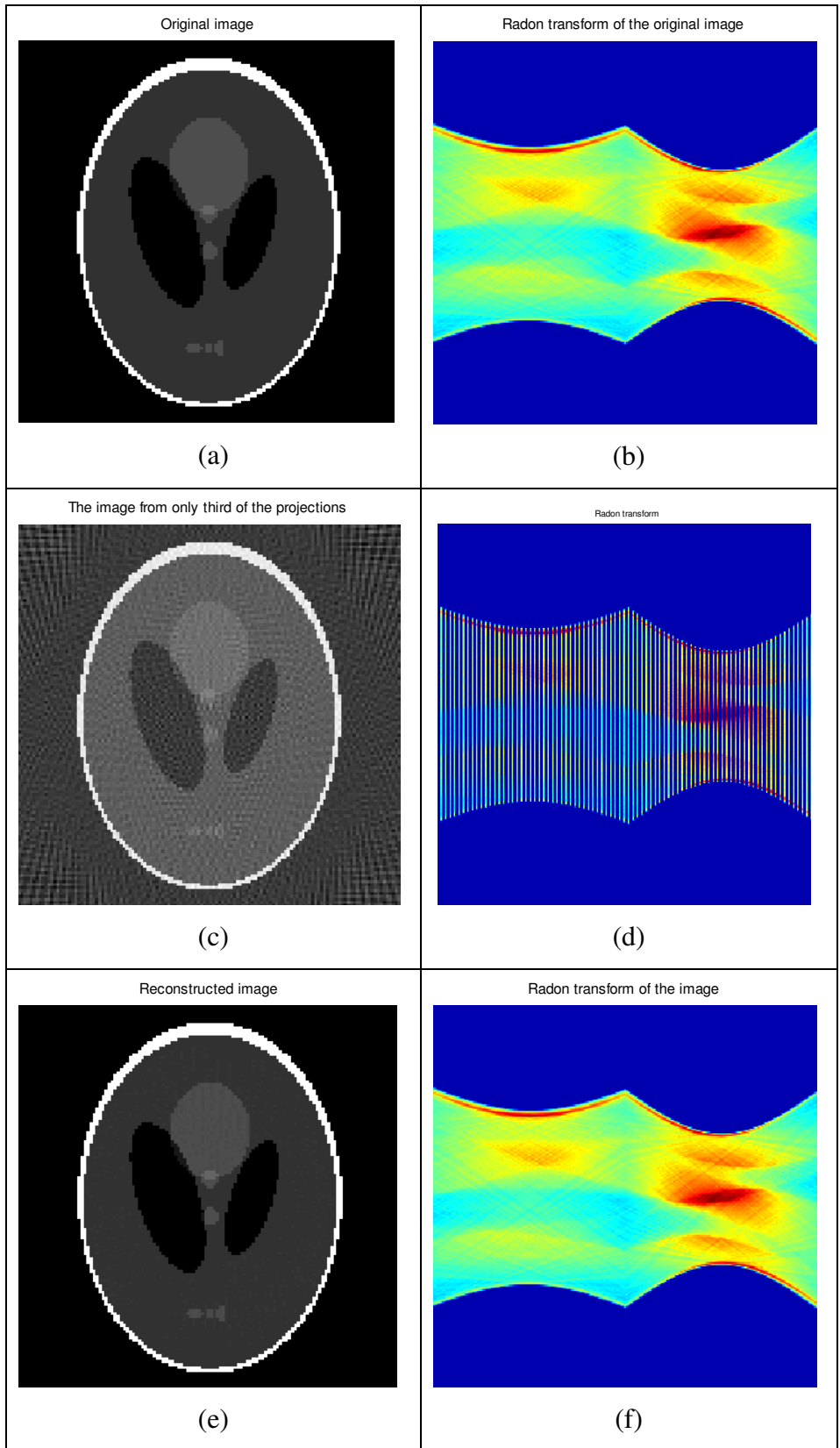

**Figure 5-7: (a)-(b) Original CT image and its Radon transform, (c)-(d) Original CT image after removing every second and third column of its sinogram. (e)-(f) Reconstructed CT image**

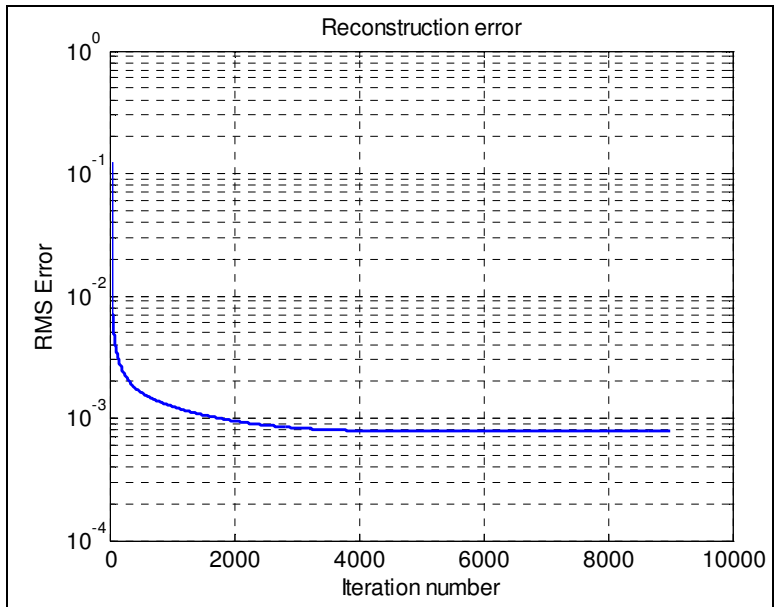
**Figure 5-8: RMS error of the reconstruction algorithm**

As can be seen, the reconstruction algorithm based on the discrete sampling theory is very efficient, and good results are obtained even when significant number of the projections has been removed, when choosing the appropriate domain for reconstruction and using an a-priori knowledge about the finite support of the image.

# 6  SUMMARY AND CONCLUSION

In the first part of this thesis, a computer model for studying the performance of the algorithm for generating super-resolved images from sequences of low resolution images distorted by turbulences was described, as well as the obtained results of the study.

The study has shown that:
- Image local instabilities in video sequences distorted by atmospheric turbulence can be compensated and utilized for increasing image resolution beyond the limits defined by the camera sampling rate.
- Camera fill factor limits potential image resolution enhancement that could be achieved by means of fusing several low-resolution images. Cameras with smaller fill factor are better suited for the super resolution process. In this respect, color cameras with separate RGB sensors are most promising.
- Most suitable for super resolution are turbulent videos in which standard deviation of pixel displacement due to turbulence is of the order of 0.5 inter-pixels distance.
- Several tens of image frames with low inter-frame correlations of pixel displacements are required to achieve substantial resolution enhancement.
- Good re-interpolation of images fused from a set of low-resolution turbulent images is essential for achieving good quality of resolution enhancement.

The second part of this research, dealt with reconstruction of signals from sparse data. This is the last step in the super resolution algorithm, but in a wider perspective, it is quite a common problem in many other applications. It was shown, that since any discrete signal can be expressed as a linear combination of basis functions of a certain transform, the problem of signal recovery from sparse data can be formulated as a linear algebra one of solving a set of linear equations by means of matrix inversion. The rank of the matrix defines the effective number of samples.  The feasibility of solving this set of equations depends on the transform selected for signal representation, and on the locations of the samples.  This approach is based on the formulated discrete sampling theorem and on the notion of discrete signal "band limitation" in terms of their representation in a domain of a certain transform.

Since there is a virtually infinite number of signal transform domain representations, one needs, in order to secure the best signal approximation in terms of mean square approximation error, to find an appropriate transform that has better energy compaction for representing this signal. This is usually can be done on the base of a-priori knowledge about the signal.

As a numerical tool for signal recovery from sparse data, two algorithms were suggested; the direct matrix inversion and the iterative Gerchberg-Papoulis algorithm. The latter can be used when signal dimensions are too high for the direct matrix inversion.

Several signal discrete transforms were considered: Fourier (along with DCT), Haar, Walsh and D4-wavelet. For the classical case of Fourier, it was shown that the location of the samples is irrelevant, as long as there are enough samples. For Haar reconstruction, the reconstruction is simply nearest neighbor interpolation. It was also found that, even when perfect reconstruction is possible, the location of the samples affects the speed of convergence when using iterative reconstruction.

As an application example of the theory, CT reconstruction of images from sparse or sparsely sample projections was demonstrated, that make use of the fact that many reconstructed images have considerably large empty backgrounds and therefore their sinograms are "band limited" in the inverse-Radon domain.

## תקציר

עיוות נפוץ בסרטי וידאו הוא תנודותיות אקראית, כזו שכללית לכל התמונה, או כזו המשתנה באיזורים שונים בתמונה. העיוות נגרם, למשל, מחוסר יציבות של המצלמה או שבירה של קרני האור כתוצאה משינוי במקדם השבירה של התווך או מסיבות אחרות.

סרטים אלה, המציגים לעתים תכופות עצמים נעים על רקע קבוע, מכילים המון מידע יתיר, כזה הניתן לניצול על מנת לייצב את התמונה כשיש הפרדה בין האובייקט הנע לבין הסביבה הסטטית. לאחרונה, הוצע להשתמש ביתירות הזו, על מנת להעלות את הרזולוציה תוך שימוש בקביעת מיקום ברמת דיוק תת-פיקסלית לסט של כמה פריימים מהוידאו השייכים לרקע הקבוע.

מטרת העבודה שנעשתה, הנה לחקור את היכולות והמגבלות של אלגוריתמי העלאת רזולוציה מסוג זה. לחיבור זה שני חלקים: בחלק הראשון, אנו חוקרים, בעזרת הדמיית מחשב, את ההשפעה של מגוון פרמטרים על העלאת הרזולוציה הניתנת להשגה. פרמטרים כמו Fill Factor, עוצמת התנודתיות ומספר הפריימים המשמשים כקלט לאלגוריתם. חלק מרכזי בתהליך הסופר-רזולוציה, הינו שחזור האות (במקרה זה, התמונה) ממידע חלקי הנצבר מהמיקום האקראי של הפיקסלים בפריימים השונים. לכן, החלק השני של העבודה עוסק בשחזור אותות ממידע חלקי. אנו מציגים את משפט הדגימה לאותות בדידים (דיסקרטים), ומראים שבהנתן חלק מהדגימות של האות הבדיד, ניתן למצוא קירוב, מוגבל סרט, במישור של התמרה כלשהי, שיהיה עם שגיאה ריבועית ממוצעת מינימלית (MMSE). אנו גם מנתחים את המגבלות הקשורות למיקום הדגימה בייצוג ע"י פונקציות בסיס שונות על מנת לקבל שחזור אופטימלי, ומראים שבמקרה בו משחזרים אות מוגבל סרט (כך שהאות מכיל תדרים נמוכים עד תדר מסויים) במישור פוריה (פונקציות בסיס DFT), האות ניתן לשחזור מדוייק, ללא קשר למיקום הדגימות. לאחר מכן התוצאות מורחבות לשחזור תמונה הנוצרה מהיטלים (כגון תמונות CT) כאשר חלק מהיטלים חסרים או כאשר חלק מהדגימות על חלק מהיטלים חסרים.

# יכולות ומגבלות של אלגוריתמי סופר-רזולוציה ושיחזור של אותות דגומים חלקית

## גיל שבת

אוניברסיטת תל-אביב

הפקולטה להנדסה ע״ש איבי ואלדן פליישמן

בית הספר לתארים מתקדמים ע״ש זנדמן-סליינר

# יכולות ומגבלות של אלגוריתמי סופר-רזולוציה ושיחזור של אותות דגומים חלקית

חיבור זה הוגש לקראת התואר ״מוסמך אוניברסיטה״ בהנדסת חשמל ואלקטרוניקה

על-ידי

## גיל שבת

סיון תשס״ח